\documentclass{article}

\usepackage{arxiv}

\usepackage[utf8]{inputenc} 
\usepackage[T1]{fontenc}    
\usepackage{url}            
\usepackage{booktabs}       
\usepackage{amsfonts}       
\usepackage{nicefrac}       
\usepackage{microtype}      
\usepackage{pgf,tikz}
\usetikzlibrary{arrows}
\usepackage{amsmath,amsfonts,amssymb} 
\usepackage{graphicx}	
\usepackage{caption}
\usepackage{subcaption}
\usepackage{longtable}
\usepackage{epstopdf}
\usepackage{float}
\usepackage{multirow}
\usepackage{algorithmicx}
\usepackage{algorithm}
\usepackage{algpseudocode}
\usepackage[numbers, sort]{natbib}
\usepackage{xcolor}

\DeclareMathAlphabet\mathbfcal{OMS}{cmsy}{b}{n}

\title{Heuristics for optimizing 3D mapping missions over swarm-powered ad hoc clouds}

\date{} 					

\author{ \hspace{1mm}Leandro~R.~Costa\\
	Department of Mathematical and Industrial Engineering\\
	Polytechnique Montréal\\
	Montreal, Canada \\
	\texttt{leandro.costa@polymtl.ca} \\
	\And
	\hspace{1mm}Daniel~Aloise \\
	Department of Computer and Software Engineering\\
	Polytechnique Montréal\\
	Montreal, Canada \\
	\texttt{daniel.aloise@polymtl.ca} \\
	\And
	\hspace{2.5cm}Luca~G.~Gianoli \\
	\hspace{2.5cm}Humanitas Solutions\\
	\hspace{2.5cm}Montreal, Canada \\
	\hspace{2.5cm}\texttt{luca@humanitas.io} \\
	\And
	\hspace{2.1cm}Andrea~Lodi \\
	\hspace{2.1cm}Department of Mathematical and Industrial Engineering\\
	\hspace{2.1cm}Polytechnique Montréal\\
	\hspace{2.1cm}Montreal, Canada \\
	\hspace{2.1cm}\texttt{andrea.lodi@polymtl.ca} \\
}

\renewcommand{\undertitle}{}

\begin{document}
\maketitle

\begin{abstract}
Drones have been getting more and more popular in many economy sectors. 
Both scientific and industrial communities aim at making the impact of drones even more disruptive by empowering collaborative autonomous behaviors --- also known as swarming behaviors --- within fleets of multiple drones.
In swarming-powered 3D mapping missions, unmanned aerial vehicles typically collect the aerial pictures of the target area whereas the 3D reconstruction process is performed in a centralized manner.  
However, such approaches do not leverage computational and storage resources from the swarm members.
We address the optimization of a swarm-powered distributed 3D mapping mission for a real-life humanitarian emergency response application through the exploitation of a swarm-powered ad hoc cloud.
Producing the relevant 3D maps in a timely manner, even when the cloud connectivity is not available, is crucial to increase the chances of success of the operation.
In this work, we present a mathematical programming heuristic based on decomposition and a variable neighborhood search heuristic to minimize the completion time of the 3D reconstruction process necessary in such missions.
Our computational results reveal that the proposed heuristics either quickly reach optimality or improve the best known solutions for almost all tested realistic instances comprising up to 1000 images and fifteen drones.
\end{abstract}


\section{Introduction}
\label{sec:introduction}

Unmanned Aerial Vehicles (UAVs), which are also referred to as drones, are a remotely operated aircraft.
Their aerial capabilities and low cost made them an attractive option for operations as building inspection, photo collection, and area surveillance.
That explains their popularity and adoption in a multitude of sectors of the economy.

Besides being remotely operated by pilots, drones can also operate autonomously when obeying an on-board flight controller.
This is handled by employing a collaborative intelligence program on each agent of the fleet while keeping them connected on the same wireless network.
This  collaborative capability that simulates natural swarms~\cite{tan2013research} (e.g., ant colonies, bee swarms, and bird flocks)  is further leveraged by applying swarm robotics to conceive a fleet of fully autonomous drones focusing on fulfilling a common mission.
Swarm robotics studies how to coordinate, in a distributed and decentralized manner, a large group of simple embodied robots to perform collective tasks and maximize the swarm performance~\cite{csahin2004swarm}. 
Such swarm behavior is a powerful tool to foster UAV applications, given its capability to deliver both performance and resilience without requiring any centralized control.
Swarming UAVs are deployed to support operations in a long list of domains~\cite{nex2014uav}, including forestry~\cite{berni2009thermal, grenzdorffer2008photogrammetric, martinez2006experimental}, archaeology and architecture~\cite{chiabrando2011uav, lambers2007combining, oczipka2009small, rinaudo2012archaeological, verhoeven2009providing}, environmental monitoring~\cite{hartmann2012determination, manyoky2011unmanned, niethammer2010uav, smith2009volcano, zhang2008uav}, emergency management~\cite{chou2010disaster, haarbrink2006helicopter, molina2012searching} and precision agriculture~\cite{zarco_tejada2014, bendig2014, diaz_varela2015}.

Likewise, the swarm robotics has drawn the attention of the operations research community as an attractive opportunity to improve the efficiency of swarm-powered missions~\cite{alena2018optm_uav, coutinho2018}.
For instance, decentralized optimization methods have been leveraging search problems~\cite{Sujit2004search, oh2014road_search, oh2015road_relay, lanillos2014, gan2011multi, ji2015cooperative}, target assignment problems~\cite{Tang2005MMMT, KARAMAN2008CG, niccolini2010multiple, viguria2010distributed, choi2011genetic, barrientos2011aerial, Moon2013, moon2015, turpin2014capt,enright2015handbook, sadeghi2017heterogeneous}, node covering problems~\cite{ZORBAS201616, Ladosz2018}, scheduling problems~\cite{CARABALLO2017725}, etc.

UAV swarming solutions are typically used in synergy with other technologies such as digital photogrammetry~\cite{Gomarasca2009, Linder2016}, which focuses on extracting and displaying the relevant 2D/3D geometric information from the portrayed physical environment.
Given a set of pictures that are fairly distributed across the area of interest --- multiple shooting points and shooting perspectives should be considered for improved performances --- it is possible to extract a 3D model of the region itself by performing a so-called \emph{three-dimensional reconstruction}.
In terms of synergy the between swarming and photogrammetry technologies, the 3D reconstruction process is typically included in a swarm-powered mission pipeline where swarming UAVs are responsible for collecting, as fast as possible, the set of aerial images required to properly build the 3D map of the desired area.

Although the aerial photo collection is typically carried out in parallel by multiple agents, current 3D mapping literature addresses the 3D-reconstruction process in a centralized manner~\cite{doherty2016collaborative, loianno2016swarm, schmiemann2017distributed, kobayashi2017method, golodetz2018collaborative, milani2016impact}. 
\citet{meyer2015optimizing} employed 3D reconstruction solutions which were conducted in a centralized base-station to survey a heritage site in Mexico, while other authors focused on multi-sensor data fusion to feed the 3D reconstruction algorithms with more accurate data~\cite{hinzmann2016collaborative, surmann20173d}.

Such approaches are susceptible to internet connectivity and network latency issues.
Thus, the distributed power within the UAV swarm can be exploited establishing an ad hoc cloud infrastructure able to safely perform the 3D reconstruction of the considered 3D mapping mission~\cite{chen2019,hamdaqa2015,malhotra2014,yaqoob2016}. 
This strategy profits from the computational power in the microcomputers installed on the swarming UAVs instead of the computational resources present in a powerful computer (e.g.~\cite{cloudODM}).

The motivation of this paper arises from a real-life problem in the emergency response field. 
The decision-making and situation awareness during a first response operation are highly boosted by 3D maps of the affected region.  
Such 3D models allow the first responders to detect relevant threats like damage  in  roads  and  buildings  or  insecure  zones.
Thus, creating 3D maps must be quickly done regardless internet connectivity to enhance the safety and efficiency of the first responders.

In~\citet{capsac2020}, the authors introduced the Covering-Assignment Problem for swarm-powered ad hoc clouds (CAPsac), whose objective, in the context of a 3D mapping UAV mission, is to optimally generate and place the multi-node computing workload that will be responsible for performing the 3D reconstruction process. 
Since each computing node is responsible for reconstructing a specific sub-region on a specific drone, the optimal solution of the problem describes how to minimize the processing time by optimally i) splitting the set of available photos to form multiple sub-regions and then ii) assigning each sub-region to a specific UAV.

During emergency field operations, each minute counts and nearly optimal solutions of the CAPsac problem must be computed as quickly as possible. 
In this way, 3D maps can be promptly put in the hands of the emergency responders, while leaving the UAVs available to perform additional critical tasks --- including other 3D mapping missions. 
Furthermore, a fast CAPsac solution allows the flying UAVs to preserve the battery life by limiting the idle flying periods spent waiting for 3D processing instructions. 
Note that the autonomy of typical commercial drones does not go beyond one hour.

In this paper, we propose a Variable Neighborhood Search (VNS) heuristic~\cite{pierre_nenad2001vns} based on \textit{sub-tree reconstruction}, \textit{splitting hyperplane reallocation}, \textit{sub-region transfer}, and \textit{sub-region swap} neighborhoods for quickly optimizing a swarm-powered 3D mapping mission according to CAPsac. 
We also develop a mathematical programming-based heuristic, namely \emph{Decomposition-based heuristic}, to assess the performance of matheuristics.

The paper is organized as follows.
The next section describes how to cast a swarm-powered 3D mapping mission as CAPsac problem. 
The proposed mathematical programming-based heuristic is presented in Section~\ref{sec:decomp_heur}.
Likewise, the VNS fundamentals and the proposed VNS-based heuristic are exposed in Section~\ref{sec:vns_for_CAPsac}. 
Finally, Section~\ref{sec:experiments} shows and analyzes the computational results obtained over realistic instances, while Section~\ref{sec:conclusions} outlines our concluding remarks.

\section{Swarm-powered 3D mapping mission as a CAPsac}
\label{sec:CAPsac}

A swarm-powered 3D mapping mission is decomposed in two main phases~\cite{capsac2020}:
\begin{enumerate}

    \item \textbf{Photo collection}: the UAVs of the swarm dynamically collaborate to collect all the necessary aerial pictures of the area of interest. 
Note that the set of required pictures is typically computed by dedicated mapping software and is merely an input of the mapping mission~\cite{pepe2018planning}.

    \item \textbf{3D processing}~\cite{tang2020}: the collected pictures are collaboratively processed by the microcomputers installed on the UAVs to produce a 3D map.
During this process, the computing workload can be parallelized over the available computing units. 
Furthermore, pictures can be transferred over the inter-drone wireless network to satisfy the input requirements of the distributed reconstruction tasks.

\end{enumerate}

\begin{figure}[H]
    \centering
    \input{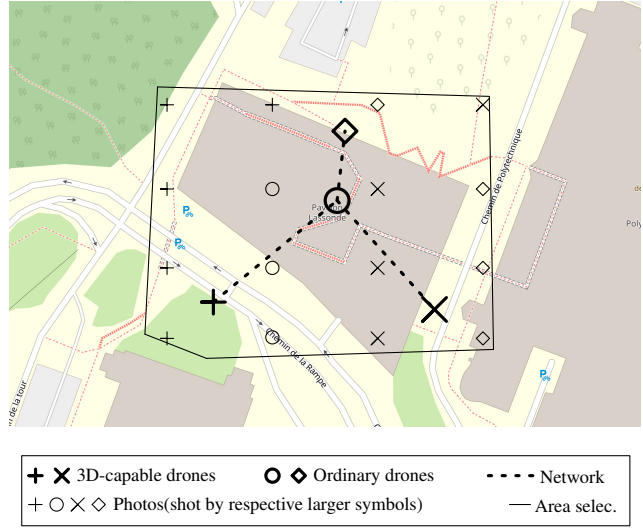}
    \caption{Swarm-powered 3D mapping mission as an instance of CAPsac. ©2020 IEEE~\cite{capsac2020}}
   	\label{fig:intro_entry}
\end{figure}

By casting a swarm-powered 3D mapping mission as a CAPsac, we aim to optimize the 3D processing phase only, with no direct control over the photo collection step. 
Therefore, we consider that the set of photos taken by the drones as well as their locations are input parameters for CAPsac solution. 
A swarm-powered 3D processing application employing a swarm of four UAVs is illustrated in Figure~\ref{fig:intro_entry}~\cite{capsac2020}.

In that example, the large “$\times$”, “$+$”, “$\diamond$”, and “$\circ$” identify the drones.
Only drones with powerful microcomputers able to sustain the 3D reconstruction methods are considered \emph{3D-capable}~\cite{capsac2020}.
The 3D-capable drones are represented by the $+$ and the $\times$ symbols. 
That is, $+$ and $\times$ are responsible for performing the 3D processing stage.
The region portrayed by the set of photos $P$ is bounded by the continuous lines.
Further, the photos are represented by the small “$\times$”, “$+$”, “$\diamond$”, and “$\circ$”.
Their symbols match with the UAV where the photo is stored, for instance, pictures identified by a small $+$ are stored in the UAV represented by the large $+$.

A solution to the CAPsac problem describes how to parallelize optimally a massive 3D reconstruction task into smaller 3D reconstruction sub-tasks to be distributed across the 3D-capable drones. 
This means that a 3D reconstruction sub-task is associated with a specific sub-region of the target region.
Therefore, the optimal solution of CAPsac minimizes the completion time (i.e., makespan) of the whole 3D reconstruction phase.
When optimizing the CAPsac, three constraints cannot be neglected~\cite{capsac2020}: 
the 3D reconstruction sub-regions must be a \emph{spatial-convex covering}; 
the communication delays follow the \emph{Max-Min Fairness} (MMF) paradigm~\cite{bertsekas1992data}; 
the 3D reconstruction sub-tasks are distributed across the 3D-capable drones according to a \emph{reliability factor}.

More precisely, the sub-regions form a \emph{spatial-convex covering} if and only if the union of all sub-regions (subsets of photos) is equal to the target region and all sub-regions are a \emph{spatial-convex set} --- all photos lying inside the set's convex hull are allocated to that same sub-region~\cite{capsac2020}.
Figure~\ref{fig:realistic_good_set} shows an example of a spatial-convex set, whereas Figure~\ref{fig:realistic_bad_set} illustrates a set of photos which is \emph{not} a spatial-convex set.
In those figures, photos allocated to the set are represented by the filled “$\bullet$” and the empty “$\circ$” represent photos not allocated to the set.

\begin{figure}[H]
    \centering
    \begin{minipage}{0.45\textwidth}
        \centering
		\input{convex_set_figure.tex}
		\caption{Spatial-convex set and the respective convex hull. Adapted ©2020 IEEE~\cite{capsac2020}}
		\label{fig:realistic_good_set}       
    \end{minipage}
	\hspace{5mm}\begin{minipage}{0.45\textwidth}
	    \centering
		\input{non_convex_set_figure.tex}
		\caption{Ordinary set and the respective convex hull. ©2020 IEEE~\cite{capsac2020}}
		\label{fig:realistic_bad_set}         
    \end{minipage}
\end{figure}  

The communication delays must be taken into account since the photo collection stage is not optimized in the CAPsac, and the 3D reconstruction of a sub-region cannot start until its assigned drone has in its hard disk all photos belonging to that sub-region, that is, all 3D reconstruction input photos.
Thus, a drone must request to the other drones the photos to complete its designed sub-region. 
A single tree network topology (dotted lines in Figure~\ref{fig:intro_entry}), on top of which establishing TCP sessions --- one per photo transfer, is adopted as inter-drone communication means. 
The MMF paradigm is considered to approximate the TCP-based rate allocation for multiple photo transfer sessions occurring within the swarm communication tree~\cite{massoulie1999bandwidth}.

\begin{figure}[H]
    \centering
	\input{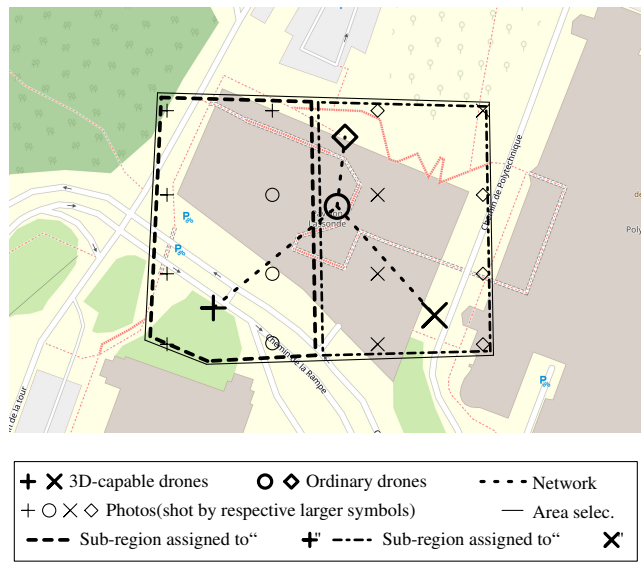}
   	\caption{Spatial-convex covering and its assignment optimizing the makespan of a 3D mapping mission.©2020 IEEE~\cite{capsac2020}}
    \label{fig:intro_result}
\end{figure}

Finally, the sub-region (3D reconstruction sub-task) to drone assignments must be robust to drone malfunction. 
This is addressed by introducing a \emph{reliability factor}, which dictates the minimal number of drones each sub-region must be assigned.
For instance, a solution for a reliability factor equal to one is illustrated in Figure~\ref{fig:intro_result}.
There, the two spatial-convex sets (the same number as the 3D-capable drones) are represented by the dashed (left) and the dashed-and-dotted lines (right).
The sub-region on the left is assigned to the UAV “$+$” and the sub-region on the right is assigned to the UAV “$\times$”.

Let us now define the notation for CAPsac. 
We consider the region photographed by the set of photos $P$ as the region targeted by the 3D processing stage. 
The CAPsac considers that photo positioning was performed beforehand, and then, the locations of all captured photos in $P$ are fixed and known.
Further, let $\lambda_{p}$ and $\mu_p$ be the estimated photo processing time and
the data size of photo $p \in P$, respectively. 
Besides, a set of drones $D$ is at disposal but only a subset $\bar{D} \subseteq D$, where $|\bar{D}| = m$, is able to perform the 3D reconstruction.
Therefore, the number of sub-regions constructed by the CAPsac is equal to $m$.
Given a drone $d \in D$, the binary parameter $\theta_{dp}$ indicates if $d$ has the photo $p \in P$ in its hard disk.

The drones are disposed in an undirected tree topology $T = (N,A)$, in which $N$
corresponds to the set of nodes where each drone is located and $A$ comprises the sets of arcs linking the nodes in $N$. 
The set $F$ is composed by all transmission demands across pairs of drones, such that $F = \{ f^{hd} | (h,d) \in D \times \bar{D} \}$, in which a specific transmission demand from the drone $h$ to the drone $d$ is represented by $f^{hd}$. 
For each demand $f^{hd}$, let $V^{hd}$ be the set of arcs $(i,j) \in A$ in the unique routing path between $h$ and $d$, and
$\overline{F}_{ij} $ the set of demands using the arc $(i,j)$, i.e.,  $\overline{F}_{ij} = \lbrace f^ {hd} \in F | (i,j) \in V^{hd} \rbrace$.
Given an arc $(i,j) \in A$, let $c_{ij}$ be the capacity of the arc $(i,j)$.
Also, denote by $\bar{c}^{hd}$ the minimum $c_{ij}$ in $V^{hd}$. 
Finally, each transmission demand is allowed within a time limit $\hat{T}$, after which it is considered as infeasible. 

All the notation of the presented parameters is presented in Table~\ref{tab:par}.
\begin{table}[h]
	\centering
	\caption{CAPsac parameters for a 3D mapping mission. Adapted ©2020 IEEE~\cite{capsac2020}}
	\begin{tabular}{ l  l}
	\hline\noalign{\smallskip}
	Parameters & Description \\
   \noalign{\smallskip}\hline\noalign{\smallskip} 
	$\lambda_{p}$ & estimated processing time of photo $p$\\
	$\mu_{p}$ & amount of data of photo $p$ \\
	$\theta_{dp}$ & equal to 1 if drone $d$ has the photo $p$ stored in its memory \\
	$F$ & set of traffic demands between each pair of drones \\
	$V^{ab}$ & routing path of a demand $f^{ab} \in F$ from the drone $a$ to the drone $b$ \\
	$c_{ij}$ & transmission capacity of the link $(i,j) \in A$ \\
	$\bar{c}^{ab}$ & minimum $c_{ij}$ for $(i,j) \in V^{ab}$ \\
	$\overline{F}_{ij}$ & set of demands that use  link $(i,j) \in A$ \\	
	$\sigma$ & reliability factor \\
	$m$ & number of 3D-capable drones (equiv. number of sub-regions) \\ 
	$\hat{T}$ & maximum allowed time for transmitting photos between drones \\
    \noalign{\smallskip}\hline
	\end{tabular}
	\label{tab:par}
\end{table}

\subsection{Mathematical Formulation}
\label{sec:math_formulation}

Mathematical formulations to solve the $CAPsac$ were proposed in~\citet{capsac2020}. 
In this work, we adopt the region-based formulation ($rCAPsac$)~\cite{capsac2020}.
The $rCAPsac$ formulation minimizes the completion time of the 3D processing stage $T_{\max}$, and it exploits the set $\mathcal{S}$ containing all rectangular spatial-convex sets of photos in $P$. 
Given a photo $p \in P$, the set $\mathcal{S}_{p}$ comprises all sub-regions $S \in \mathcal{S}$ which contain photo $p$. 
The binary variables $q^{S}_{d}$ indicate if $S \in \mathcal{S}$ is selected and its 3D reconstruction is allocated to the drone $d \in \bar{D}$.
Remark that all selected sub-regions must be assigned to at least $\sigma$ (reliability factor) drones.
Also, for each $S \in \mathcal{S}$, the binary variable $o^{S}$ indicates (i.e., $o^{S}=1$) if the sub-region $S$ is used in the solution;
the $t^S$ represents the time required to perform the 3D reconstruction of $S$;
the $\mu^{hd}_S$ expresses the amount of data added into the transmission demand $f^{hd}$ when $S$ is selected. 

Since 3D reconstruction cannot start until a drone has all its input pictures, the communication time required to exchange photos among the drones cannot be neglected.
For each $f^{hd} \in F$, we denote by $z^{hd}$ the binary variable which indicates when the demand $f^{hd}$ is active --- if there exists any data to be exchanged through $f^{hd}$.
Accordingly, continuous variables $\phi^{hd}$ correspond to the transmission rate performed by the demand $f^{hd}$.

As mentioned, in the case of the CAPsac, the transmission rate allocation follows the MMF paradigm.
Such rate allocation attends the MMF if and only if there is at least one bottleneck link $(i,j) \in A$ on the routing path $V^{hd}$ of each active demand $f^{hd} \in F$~\cite{nace2008max}.
Furthermore, a link $(i,j)$ is a bottleneck of the demand $f^{hd}$ if and only if~\cite{nace2008max}
\begin{enumerate}
    \item[(i)] its capacity is saturated, i.e., $\sum_{f^{ab} \in \overline{F}_{ij}}{\phi^{ab}} = c_{ij}$ and
    \item[(ii)] the transmission rate $\phi^{hd}$ of traffic demand $f^{hd}$ is the highest among the traffic demands routed over link $(i,j)$, i.e., $\phi^{hd} \geq \phi^{ab} \quad \forall f^{ab} \in \overline{F}_{ij}$.
\end{enumerate}
Given a demand $f^{hd}$, let $w^{hd}_{ij}$ be the binary variable equal to 1 if the link $(i,j)$ is a bottleneck of $f^{hd}$.
We denote by $u_{ij}$ the highest transmission rate among the traffic demands carried by the link $(i,j) \in A$,  that is, $u_{ij} = \max\limits_{f^{ab} \in \overline{F}_{ij}} \{ \phi^{ab}\}$. 

The $rCAPsac$ formulation of the $CAPsac$ is described by the following MILP~\cite{capsac2020}.
\begin{align}
    \min\limits_{q, o} &\quad T_{\max} \label{eq:scf_obj_func} \\
    \text{s.t. } &T_{\max} \geq \sum_{S \in \mathcal{S}} t^S q^{S}_{d}  &\forall d \in \bar{D} \label{eq:scf_1}\\
    &\hat{T} \cdot \phi^{hd} \geq \sum_{S \in \mathcal{S}} \mu^{hd}_{S} q^{S}_{d} &\forall f^{hd} \in F \label{eq:scf_2}\\
	&\sum_{d \in \bar{D}} q^{S}_{d} \geq \sigma o^{S} &\forall S \in \mathcal{S} \label{eq:scf_3}\\
	&\sum_{S \in \mathcal{S}_{p}} o^{S} \geq 1 &\forall p \in P \label{eq:scf_5}\\
	&\sum_{S \in \mathcal{S}} o^{S} = m &\label{eq:scf_52}\\
	&z^{hd} \leq \sum_{S \in \mathcal{S}} \mu^{hd}_{S} q^{S}_{d} &\forall f^{hd} \in F \label{eq:scf_6}\\
	& \phi^{hd} \leq \bar{c}^{hd} z^{hd} & \forall f^{hd} \in F \label{eq:nlm_8}\\
	&\sum_{(i,j) \in V^{hd}} w^{hd}_{ij} \geq z^{hd} &\forall f^{hd} \in F \label{eq:nlm_9} \\
	&\sum_{f^{ab} \in \overline{F}_{ij} } \phi^{ab} \leq c_{ij} &\forall (i,j) \in A \label{eq:nlm_10} \\
	&\sum_{f^{ab} \in  \overline{F}_{ij} } \phi^{ab} \geq c_{ij} w^{hd}_{ij} &\forall (i,j) \in A, \forall f^{hd} \in \overline{F}_{ij} \label{eq:nlm_11} \\
	&u_{ij} \geq \phi^{hd} &\forall (i,j) \in A, \forall f^{hd} \in  \overline{F}_{ij} \label{eq:nlm_12} \\
	&\phi^{hd} \geq u_{ij} - c_{ij}(1 - w^{hd}_{ij}) & \forall (i,j) \in A, \forall f^{hd} \in  \overline{F}_{ij} \label{eq:nlm_13} \\
	&o^{S}, q^{S}_{d} \in \{ 0, 1 \} &\forall S \in \mathcal{S}, \forall d \in \bar{D} \label{eq:scf_7} \\
	&w^{hd}_{ij} \in \{0, 1\} &\forall (i,j) \in A, \forall f^{hd} \in F \label{eq:scf_8} \\
	&\phi^{hd} \geq 0, z^{hd} \in \{0, 1\} &\forall f^{hd} \in F \label{eq:scf_9} \\
	&u_{ij} \geq 0 &\forall (i,j) \in A \label{eq:scf_10}.
\end{align}
The objective function (\ref{eq:scf_obj_func}) minimizes the makespan of the 3D processing phase, i.e., the maximum processing time $T_{\max}$ across all 3D-capable drones in the swarm.
Accordingly, $T_{\max}$ is computed by the constraints (\ref{eq:scf_1}), in which, the summation $\sum_{S \in \mathcal{S}} t^S q^{S}_{d}$ determines the processing time of a given drone $d \in \bar{D}$.
Constraints (\ref{eq:scf_2}) restrict the latency of all transmission demands $F$ up to $\hat{T}$. 
Constraints (\ref{eq:scf_3}) enforce the reliability factor over all selected sub-regions.
That is, when a sub-region $S$ is selected, it is assigned to at least $\sigma$ 3D-capable drones. 
The photos $p \in P$ are always covered given the constraints (\ref{eq:scf_5}).
The constraint (\ref{eq:scf_52}) establishes $m$ (number of 3D-capable drones) as the number of sub-regions being selected.
The MMF transmission rate allocation is computed by the constraints (\ref{eq:scf_6})-(\ref{eq:nlm_13}). 
Constraints (\ref{eq:scf_6}) allow a transmission to be active only if there exist data to be sent from  drone $h$ to  drone $d$.
Constraints (\ref{eq:nlm_8}) set the variables $\phi^{hd}$ to 0 when $f^{hd}$ is inactive.
All active demands have at least one bottleneck link in their routing path according to inequalities (\ref{eq:nlm_9}).
Constraints (\ref{eq:nlm_10}) ensure that all link capacities are respected. 
Constraints~(\ref{eq:nlm_10}) and (\ref{eq:nlm_11}) jointly force that all bottleneck links will be saturated. 
That covers the first condition to a link $(i,j) \in A$ to be a bottleneck link.
The second bottleneck link condition is ensured by constraints (\ref{eq:nlm_12}) and (\ref{eq:nlm_13}).
Constraints~(\ref{eq:nlm_12}) make $u_{ij}$ greater or equal to the highest transmission rate passing through the link $(i,j)$. 
If $(i,j) \in A$ is a bottleneck link of the demand $f^{hd}$, constraints (\ref{eq:nlm_13}) ensure that $\phi^{hd}$ will not be exceeded by any other transmission rate passing thought link $(i,j)$.
The domain constraints are (\ref{eq:scf_7})-(\ref{eq:scf_10}). The number of variables of the $rCAPsac$ formulation can rapidly increase even though it is polynomial bounded by $O(|P|^4)$.

The CAPsac was proved NP-hard in~\citet{capsac2020}. A straightforward lower bound for the CAPsac is the perfect workload division among the available processing resources. 
In the context of a 3D mapping mission by UAVs, this means to equally spread the photo processing time of the whole target region by the number of 3D-capable drones. 
Given a set of photos $P$, a set of 3D-capable drones $\bar{D}$, and a reliability factor $\sigma$,  the lower bound $Lb(P, \bar{D}, \sigma)$ is given by 
\begin{equation}
Lb(P, \bar{D}, \sigma) =  \frac{ \sigma \sum_{p \in P}{\lambda_p} }{ | \bar{D} | }. 
\end{equation}

\section{Decomposition-based heuristic}
\label{sec:decomp_heur}

The formulation given in the previous section can be used to devise heuristics for the CAPsac. 
However, before that, let us define $\bar{n}$ as the perfect workload division w.r.t. the number of photos per sub-region
\begin{equation}
\bar{n} =  \frac{|P|}{| \bar{D} |}. \nonumber
\end{equation}

Inspired by the observation that good solutions have their workload w.r.t. the number of photos per sub-region close to $\bar{n}$, we decompose the problem by exploring first those solutions whose sub-regions have cardinality close to $\bar{n}$.

To that purpose, we restrict the solution space of $rCAPsac$ to sub-regions whose photo cardinalities lie inside an interval $[n_{\ell}, n_{u}  ]$, which is iteratively increased. 
That interval is initialized with $n_{\ell} = \lfloor \bar{n}  \rfloor$ and $n_u = \lceil \bar{n}  \rceil$. 
As solving this restricted $rCAPsac$ formulation leads to possibly infeasible or suboptimal solutions, one can iteratively increase the interval $[n_{\ell}, n_{u}  ]$,
and solve the new restricted CAPsac until the objective function value ($T_{\max}$) does not change between two consecutive iterations. 
This stopping condition is indeed sub-optimal. 
For example, in the presence of very contrasting photo-processing times $\lambda_p$ among the photos $p \in P$, the optimal solution might contain sub-regions with small processing times and large cardinalities.

Algorithm~\ref{alg:decomposition} summarizes the decomposition-based heuristic. 
It considers $\Omega$ the set of distinct photo cardinalities from sub-regions in $\mathcal{S}$, and a mapping function $\omega$ (implemented as a hash table) which maps the cardinalities in $\Omega$ to their respective sub-regions in $S$.
\begin{algorithm}
\caption{Decomposition-based heuristic}
\label{alg:decomposition}
\begin{algorithmic}[1]
\medskip
\State Enumerate rectangular sub-regions $\mathcal{S}$;
\State Construct and sort $\Omega$;
\State Build $\omega$ according to $\mathcal{S}$ and $\Omega$;
\State $n_{\ell} \leftarrow \lfloor \bar{n}  \rfloor$;
 $n_{u} \leftarrow \lceil \bar{n}  \rceil$;
 $\mathcal{S}^{0} \leftarrow \{ \varnothing \}$;
 $T_{\max}^{0} \leftarrow \infty$;
 $i \leftarrow 0$;
\Repeat
\State $i \leftarrow i + 1$;
\State $\mathcal{S}^{i} \leftarrow \mathcal{S}^{i-1} + \omega(n_{\ell}) + \omega(n_{u})$
\State $T_{\max}^{i} \leftarrow$ solve $rCAPsac$ encompassing sub-regions $\mathcal{S}^{i}$;
\State decrease $n_{\ell}$ to its closest smaller $n \in \Omega$;
\State increase $n_{u}$ to its closest larger $n \in \Omega$;
\Until{$T_{\max}^{i-1} = T_{\max}^{i}$} 
\emph{or} $\mathcal{S}^{i} = \mathcal{S}$
\end{algorithmic}
\end{algorithm}
Step~1 enumerates the sub-regions $\mathcal{S}$ on $O(|P|^4)$ operations~\cite{capsac2020}.
Step~2 constructs the sorted array $\Omega$ whereas step~3 creates $\omega$.
The values of $n_{\ell}$, $n_{u}$, $\mathcal{S}^{0}$, $T_{\max}^{0}$, and $i$ are properly initialized at step~4.
For each iteration (steps~5-11): 
$i$ is incremented in step~6.
Then, the set $\mathcal{S}^{i}$ is updated to include the sub-regions of cardinality equal to $n_{\ell}$ and $n_{u}$ in step~7. 
Then, step~8 solves the restricted $rCAPsac$ formulation encompassing all sub-regions in $\mathcal{S}^{i}$. 
The new interval $ [n_{\ell}, n_{u}] $ is obtained in steps~9-10 by updating $n_{\ell}$  and $n_{u}$.
The algorithm iterates (lines~5-11) until the objective function value ($T_{\max}$) does not change between two consecutive iterations, or that $\mathcal{S}^i$ contains all the sub-regions $\mathcal{S}$.
We remark that the algorithm also iterates (lines~5-11) while no feasible solutions have been found yet.
Accordingly, if the algorithm does not find a feasible solution and $\mathcal{S}^{i} = \mathcal{S}$,  the problem is proven infeasible. 
As a result, the heuristic takes at most $O(|\Omega|)$ iterations to finish in the worst case. 
In such worst scenario, the full problem (i.e., the complete $rCAPsac$ formulation) is solved once the interval $ [n_{\ell}, n_{u}  ] $ encompasses all sub-regions in $\mathcal{S}$.
It is worth pointing out that the optimal solution in one iteration remains feasible in the next one since increasing the interval $ [n_{\ell}, n_{u}  ] $ is equivalent to simply adding new columns to the $rCAPsac$ formulation.
Consequently, it can be used as a feasible upper bound solution for the next iteration.

\section{Variable Neighborhood Search for CAPsac}
\label{sec:vns_for_CAPsac}

Our VNS heuristic exploits a \textit{spatial partition tree} as a spatial data representation to ensure the spatial-convexity of sub-regions in a feasible solution. 
Furthermore, our proposed VNS is based on four neighborhoods, namely: \textit{sub-tree reconstruction}, \textit{splitting hyperplane reallocation}, \textit{sub-region transfer}, and  \textit{sub-region swap}. 
They are explained together with the VNS fundamentals and our VNS implementation in the following sections. Finally, we present how the VNS heuristic ensures communication delays feasibility during its search.

\subsection{Variable Neighborhood Search fundamentals}
\label{sec:vns}

Variable Neighborhood Search is a stochastic search metaheuristic, i.e., a local optimal evading framework based on heuristics, which aims at reaching optimal or near-optimal solutions for global combinatorial optimization problems.
VNS has been successfully applied to a vast range of NP-hard problems~\cite{hansen2019}. 

A generic combinatorial optimization problem can be formally defined as follows.
Let $N=\{1,\ldots, n \}$ be a finite set and let $c=\{c_1, \ldots, c_n\}$ be an n-dimensional vector. 
Denote by $c(F) = \sum_{i \in F}c_i$ the cost associated to the finite set $F \subseteq N$.
Given a collection of subsets $\mathcal{F}$ of $N$, a \emph{binary combinatorial optimization} problem $\mathcal{C} = (N,\mathcal{F},c)$ can be expressed as
\begin{equation}
	\min \{ c(f): f \in \mathcal{F} \}
\end{equation}

Let the n-dimensional binary vector $X^{f} = \{ x^{f}_{1},\ldots, x^{f}_{n}\}$ characterize a feasible solution such that $x^{f}_{i} = 1$ if $i \in N$ belongs to $f \in \mathcal{F}$, and  $x^{f}_{i} = 0$ otherwise. 
Thus, $\mathcal{C}$ can be seen as minimizing over a polytope, i.e.,
\begin{equation}
\min \left\lbrace c^T x | x \in conv \left\lbrace X^f \in \{0,1\}^N | f \in \mathcal{F} \right\rbrace \right\rbrace \label{comb_prob}.
\end{equation}

A local minimum $x^*$ of~(\ref{comb_prob}) is defined as 
\begin{equation}
c^Tx^* \leq c^Tx, \forall x \in \mathcal{N}(x^*), 
\end{equation}
where $\mathcal{N}(x)$ is the  \emph{neighborhood} of $x$. 
A neighborhood $\mathcal{N}(x)$ is defined as the set of neighboring solutions $x^\prime$ obtained from $x$ by systematic changes in the components of $x$ (e.g., complementing elements of $x$). 
Acknowledging that a local minimum w.r.t. a neighborhood is not necessarily the local optimum for another neighborhood,  VNS employs several different neighborhoods to escape local minima and reach global optimality.
Note that a global minimum is a local minimum regardless of the considered neighborhood.

Likewise simulated annealing and tabu search~\cite{boussaid2013survey}, VNS keeps a single solution $x$ during the whole execution of the algorithm. However, if differs from other metaheuristics mainly by its search mechanism that searches for better solutions in increasingly wider neighborhoods of $x$, according to the parameters $k \in \mathbb{Z}^{+}$ and $k_{\max} \in \mathbb{Z}^{+}$.
Let us define the neighborhoods adopted by the VNS as the set $\mathbfcal{N} = \{ \mathcal{N}_1, \ldots,\mathcal{N}_{k_{\max}} \}$ such that neighborhoods $\mathcal{N}_k \in \mathbfcal{N}$ are sorted according to their size (i.e., cardinality of $\mathcal{N}_{k}(x)$).
First,  starting with $k=1$, a random neighboring solution $x'$ is obtained from neighborhood $\mathcal{N}_k(x)$. Then, a local descent method is performed from $x'$ leading to another local minimum $x^{\prime \prime}$. 
If $x^{\prime \prime}$ is worse than (or equal to) $x$, it is dismissed and the local descent method starts from a new random neighbor $x'$ concerning the next neighborhood of $x$, i.e., $\mathcal{N}_{k+1}(x)$.
Otherwise, $x^{\prime \prime}$ replaces $x$ and the algorithm resets $k$ to 1, i.e.,  the search is resumed in the neighborhood $\mathcal{N}_1$ of the new best solution.
Every time the maximum neighborhood $\mathcal{N}_{k_{max}}$ is reached (i.e., $k=k_{\max}$), VNS restarts from the first neighborhood ($k=1$). 
The VNS iterates until the stopping condition (e.g. , maximum CPU time) is met. 
Given the order of neighborhoods in $\mathbfcal{N}$, the method favors the exploration of solutions in small neighborhoods of $x$, increasing the size of the neighborhood if necessary. 
The basic steps of VNS are given in Algorithm~\ref{alg:vns}. 

\begin{algorithm}
\caption{Variable Neighborhood Search - VNS}
\label{alg:vns}
\begin{algorithmic}[0]
\medskip
\State Select a set of neighborhoods $\mathcal{N}_k$, for $k=1,\ldots,k_{max}$;
\State Find an initial solution $x$;
\Repeat
\State $k \leftarrow 1$
\Repeat
\State (\textit{Diversification step}): Choose a random neighboring solution $x^\prime$ from $\mathcal{N}_k$;
\State (\textit{Intensification step}): Apply a local descent method from $x^{\prime}$, obtaining $x^{\prime \prime}$;
\If{$cost(x^{\prime\prime}) < cost(x)$}
  \State $x \leftarrow x^{\prime\prime}$;
  \State $ k \leftarrow 1$;
  \Else
  \State $k \leftarrow k+1$;
  \EndIf
\Until{$k > k_{max}$}
\Until{a stopping criterion is met}
\end{algorithmic}
\end{algorithm}

\subsection{Spatial partition tree}
\label{sec:spatial_part_tree}

The \emph{spatial partition tree} allows the VNS to efficiently explore the solution space without directly handling spatial-convexity constraints. 
Inspired by the \emph{k-dimensional trees} \cite{bentley1975kd_tree}, the spatial partition tree is a (\textbf{not} necessarily complete) binary tree which recursively splits the metric space under consideration.
In this kind of tree, all nodes are directly associated with a portion (sub-region) of the target region.

In the CAPsac context, the target region is discretized by the positions of the photos $P$.
Let us define $L$ as the set of distinct photo latitudes and let  $C$ be the set of distinct photo longitudes, such that $1 \leq |C| \leq |P|$ and $1 \leq |L| \leq |P|$.
The sets $C$ and $L$ offer a simple way to partition the space into rectangular sub-regions, such that
any rectangular sub-region $r$ can be represented by its left $c_{<} \in C$, right $c_{>} \in C$, inferior $\ell_{\vee} \in L$, and superior $\ell_{\wedge} \in L$ borders.
As a result, constructing a spatial partition tree splits the target region and forms a partition of rectangular sub-regions.

\begin{figure}[hp]
    \centering
   	\begin{subfigure}[b]{\textwidth}
   	    \centering
       \includegraphics[scale=0.65]{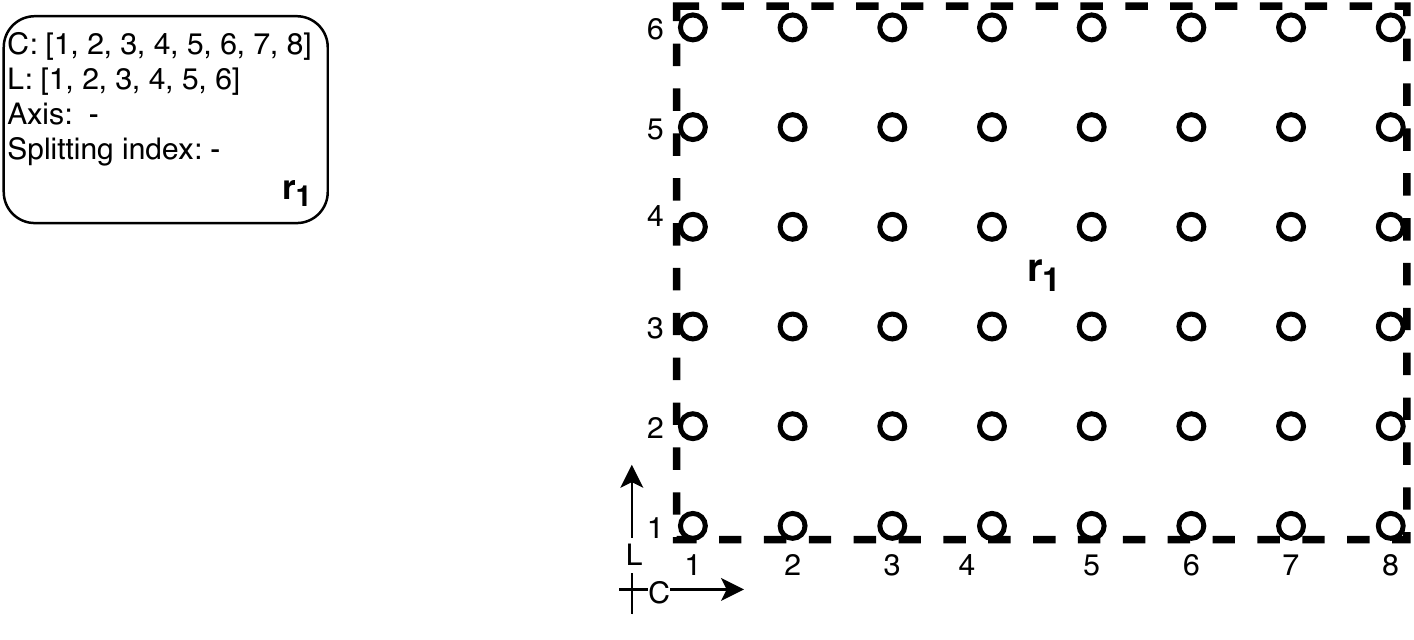}
   		\caption{Initial step. Spatial-convex tree (left) and the target region (right).}
   		\label{fig:spatial_tree_0}
    \end{subfigure}
    ~
    \begin{subfigure}[b]{\textwidth}
        \centering
       \includegraphics[scale=0.65]{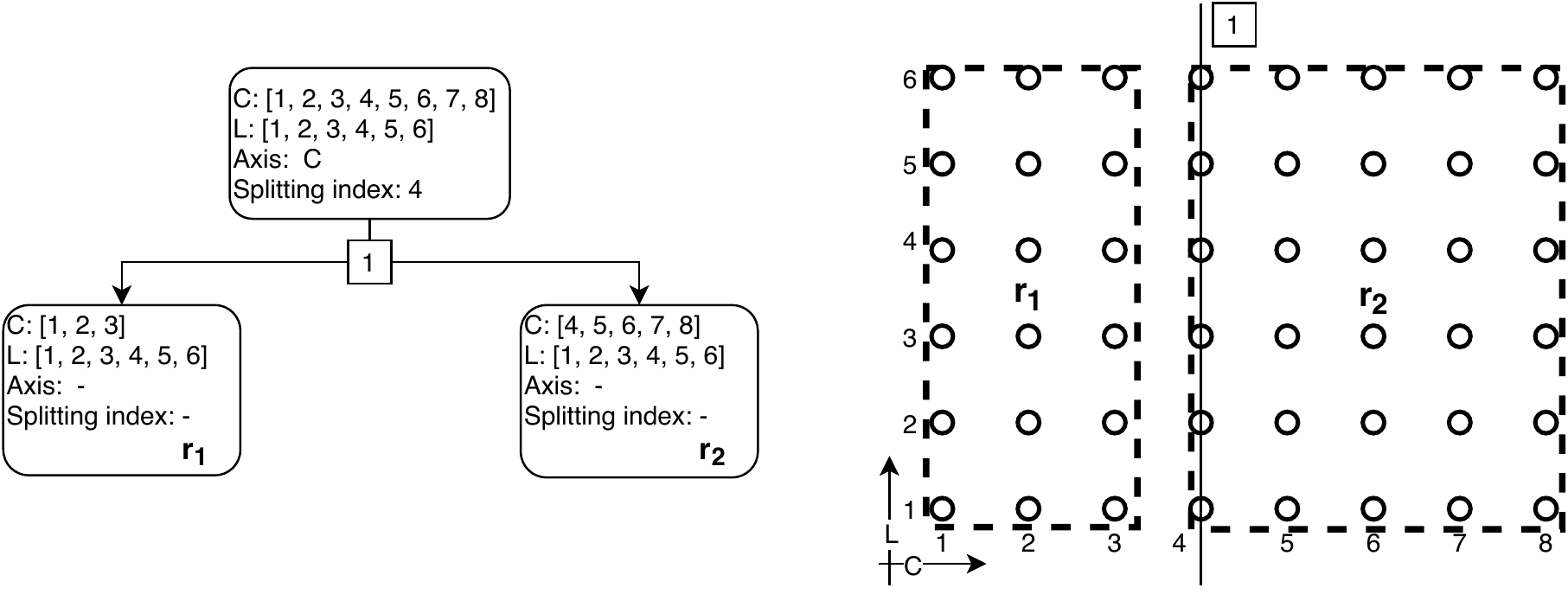}
   		\caption{Iteration 1. Spatial-convex tree (left) and its respective partition (right).}
   		\label{fig:spatial_tree_1}
    \end{subfigure}
    ~
    \begin{subfigure}[b]{\textwidth}
        \centering
       \includegraphics[scale=0.65]{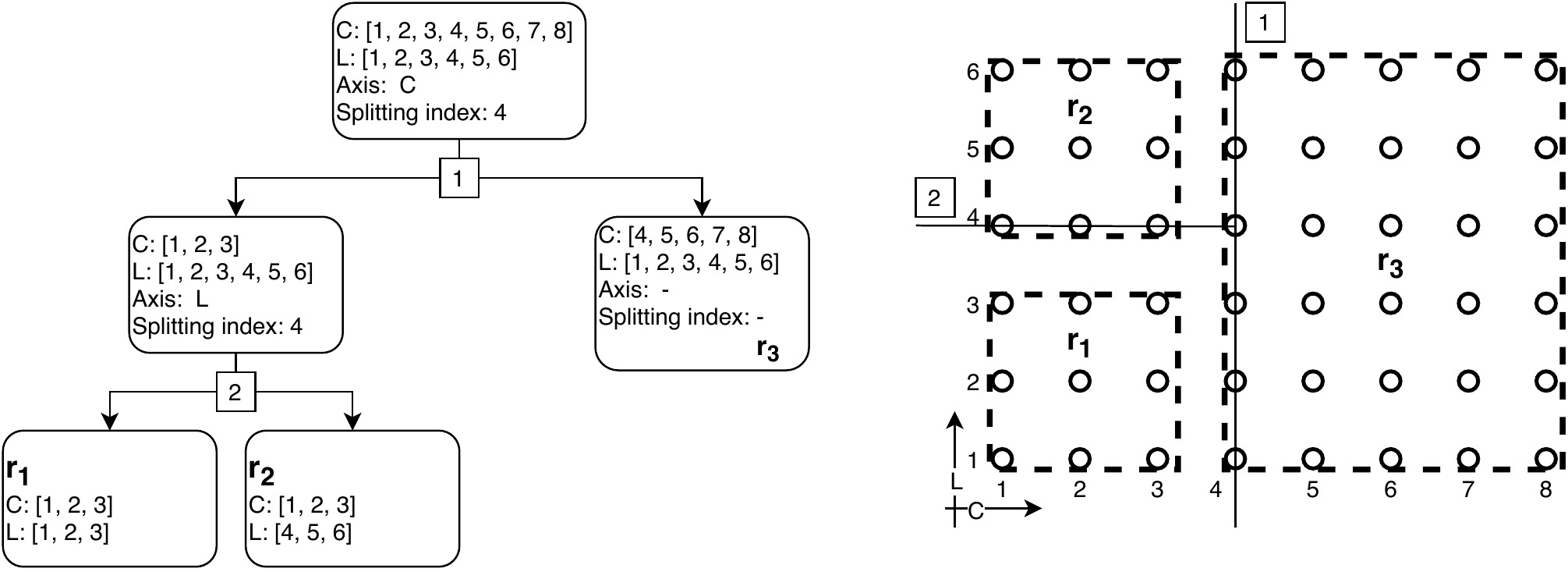}
   		\caption{Iteration 2. Spatial-convex tree (left) and its respective partition (right).}
   		\label{fig:spatial_tree_2}
    \end{subfigure}
    ~
   	\begin{subfigure}[b]{\textwidth}
   	    \centering
       \includegraphics[scale=0.7]{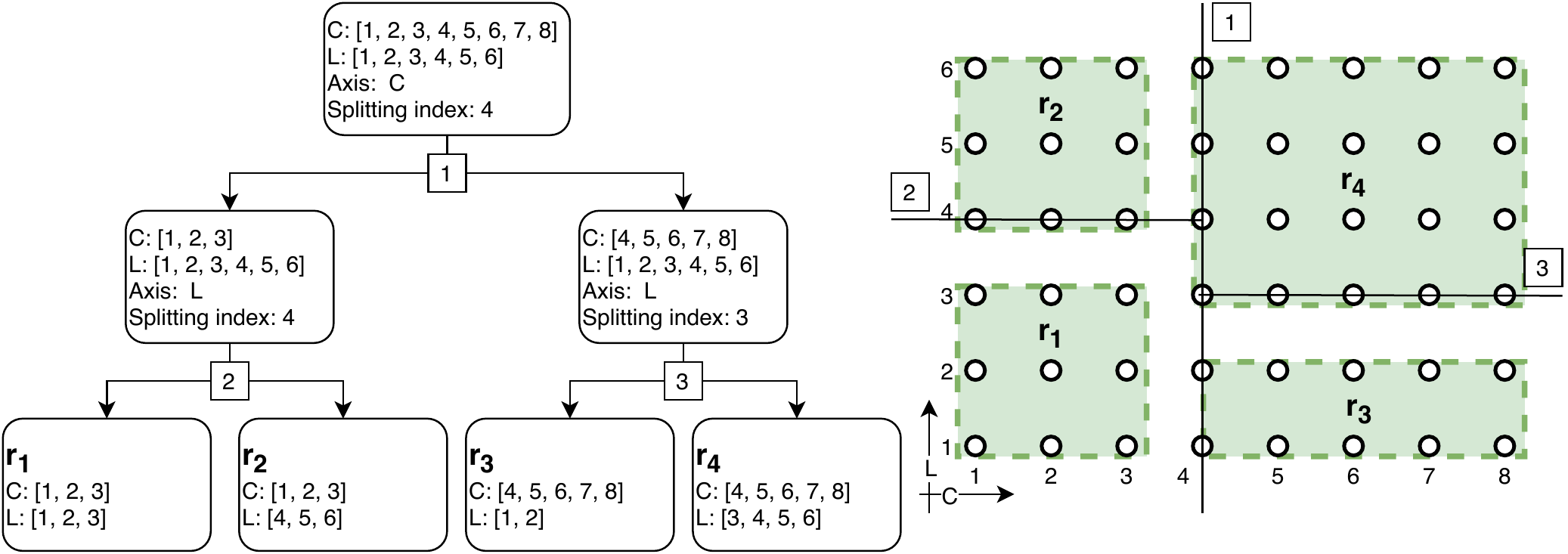}
   		\caption{Final iteration resulting in the spatial-partition tree (left) and its respective partition and splitting hyperplanes (right).}
   		\label{fig:example_spatial_tree}
    \end{subfigure}
    \caption{Illustrative construction of a spatial-convex tree with four sub-regions.}
    \label{fig:spatial_tree_contruction}
\end{figure}

A spatial partition tree is successively constructed by splitting the region of interest into new sub-regions as illustrated in Figure~\ref{fig:spatial_tree_contruction}. 
At first, the spatial partition tree contains only the root node corresponding to the whole target region as shown in Figure~\ref{fig:spatial_tree_0}.
The photos $P$ (target region), identified by the “$\circ$”, are enclosed by a unique rectangular region $r_1$ (dashed lines). 
All nodes in the tree (round-cornered rectangles) have 
(i) their associated enclosed longitudes on axis $C$ “C”); 
(ii) their associated enclosed latitudes on axis $L$ (“L”); 
(iii) the axis selected to guide the splitting (“Axis”);
(iv) the index in the selected axis chosen to define the splitting hyperplane (“Splitting index”).
It is worth to mention that the leaf nodes have the attributes (iii) and (iv) empty, and the photos inside a sub-region are implicitly given by the lists $C$ and $L$.

In each successive iteration (splitting step), a leaf node (rectangular sub-region) is selected to be split by means of an axis-aligned splitting hyperplane.
The selected splitting hyperplane (denoted by “$\square$” and represented by lines) is chosen from the location of the photos associated with a sub-region being split. 
The splitting step is repeated $|\bar{D}|-1$ times, that is, until $|\bar{D}|$ sub-regions are formed.
For instance, starting from the root node in Figure~\ref{fig:spatial_tree_0}, the construction of a partition with four sub-regions (i.e., $|\bar{D}| = 4$) is done after three iterations presented in Figures~\ref{fig:spatial_tree_1}, \ref{fig:spatial_tree_2}, and \ref{fig:example_spatial_tree}, respectively.
Finally, Figure \ref{fig:example_spatial_tree} illustrates the final spatial-partition tree and its respective partition and splitting hyperplanes, where the photos $P$ are split into the four rectangular sub-regions $r_1$, $r_2$, $r_3$, and $r_4$. 
This data structure guarantees the creation of sub-regions that are always spatial-convex sets.

\subsection{Sub-tree reconstruction neighborhood}
\label{sec:sub-tree reconstruction}

\begin{figure}[h]
    \centering
	\includegraphics[width=\textwidth]{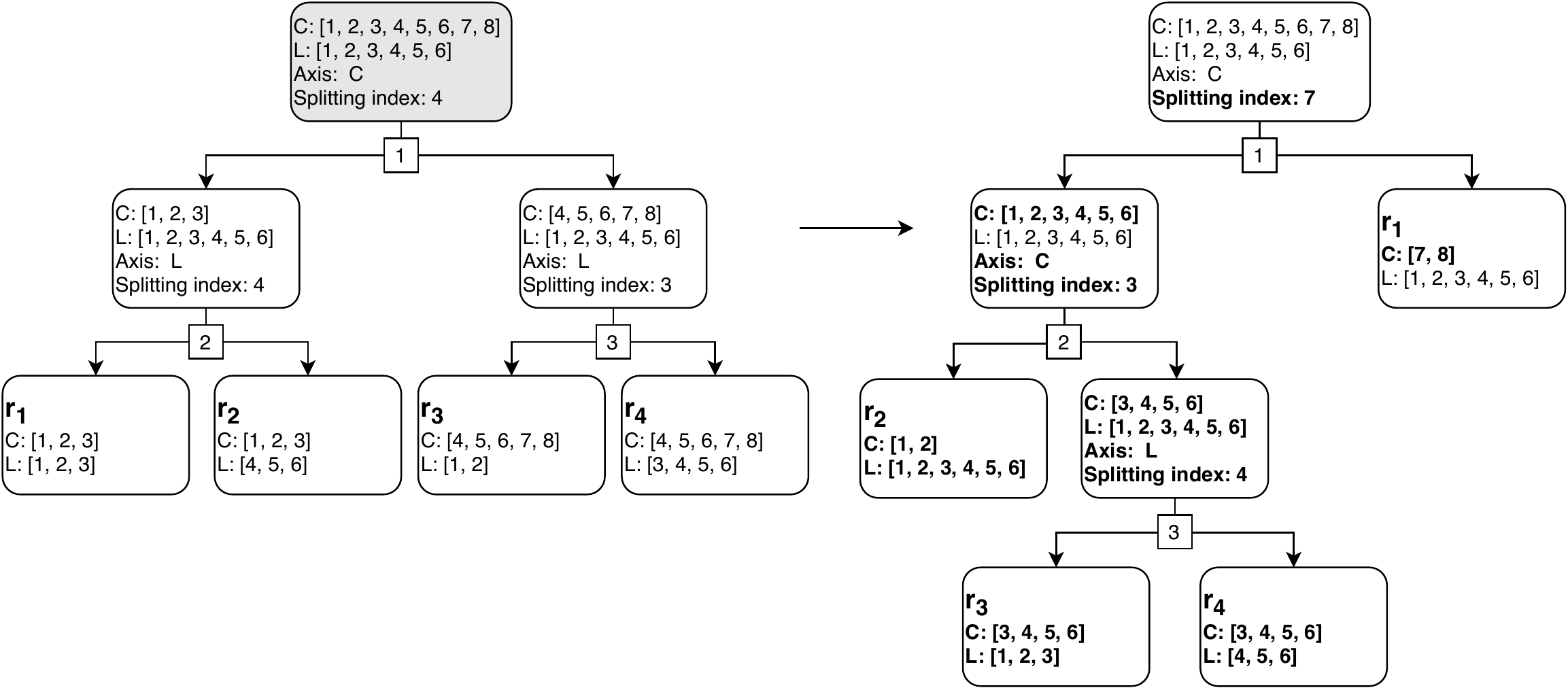}
   	\caption{Neighboring solution (right) obtained from a spatial partition tree (left) in the sub-tree reconstruction neighborhood.}
   	\label{fig:example_sub-tree_recons}
\end{figure}

Given a spatial partition tree $T$, denote by $D(T)$ the maximum depth of the nodes of $T$. 
Let $N(d)$ be the set of all nodes at the depth $d$ of $T$ where $0 \leq d \leq D(T)$.
The \textit{sub-tree reconstruction} neighborhood randomly reconstructs (as explained in Section~\ref{sec:spatial_part_tree}) a sub-tree rooted at a non-leaf node $n \in N(d)$. 
Figure~\ref{fig:example_sub-tree_recons} illustrates a neighboring solution in the right obtained from reconstructing the sub-tree rooted at the grey node (in depth 0) of the spatial partition tree in the left. 

\subsection{Splitting hyperplane reallocation neighborhood}
\label{sec:splitting_hyperplane_reallocation}

\begin{figure}[h]
    \centering
	\includegraphics[width=\textwidth]{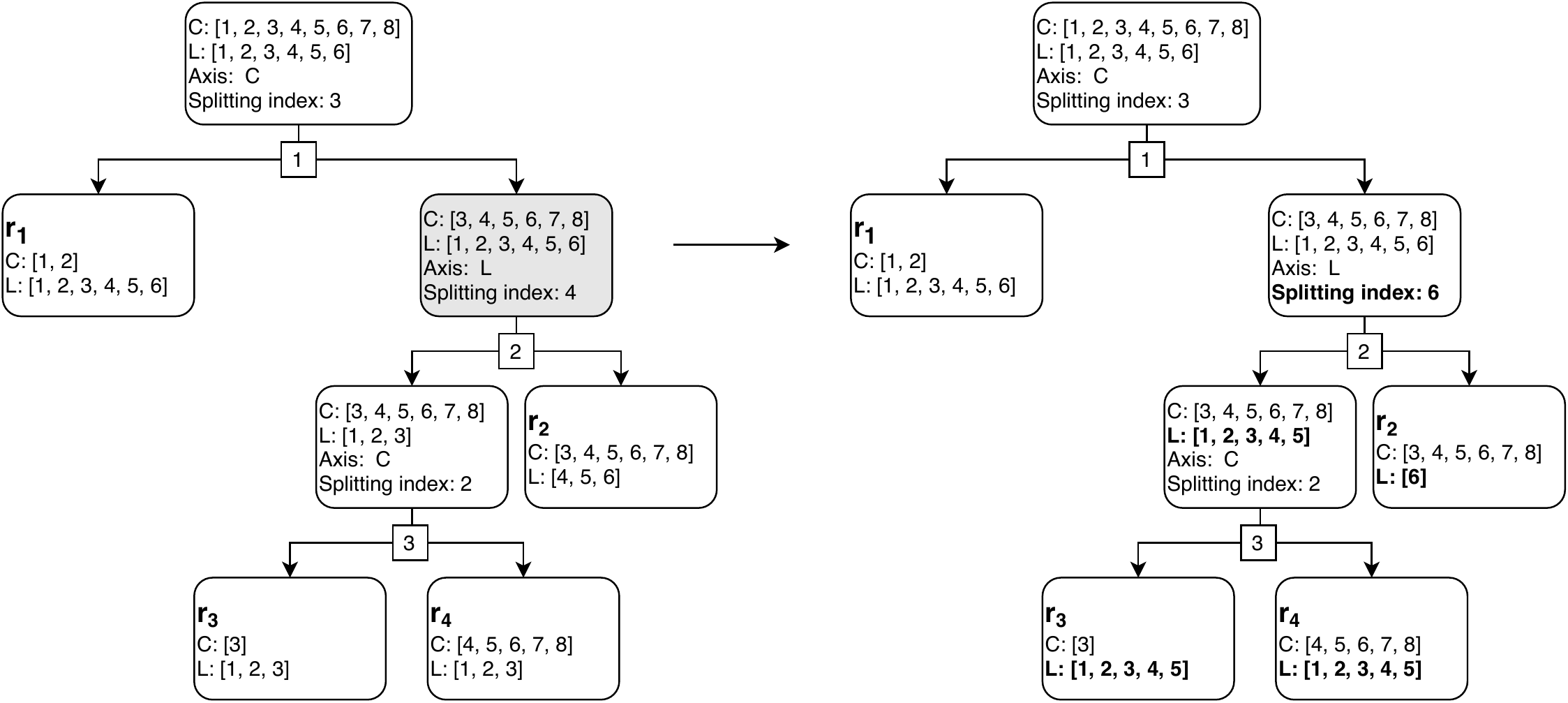}
   	\caption{Neighboring tree (right) obtained from a spatial partition tree (left) in the splitting hyperplane neighborhood.}
   	\label{fig:splitting_reallocation_example}
\end{figure}

Given a spatial partition tree $T$ and a non-leaf node $n$ which is a parent of a leaf node,
this neighborhood includes all solutions obtained by reallocating the splitting hyperplanes of the non-leaf nodes of the sub-tree rooted in $n$ (named $T_n$).
Furthermore, such reallocation is made changing the splitting index (component “Splitting index”) and/or the axis orienting the hyperplane (component “Axis”).
Unlike the sub-tree reconstruction neighborhood, the \emph{splitting hyperplane reallocation} neighborhood does not restrict $n$ to be a node in a specific depth $0 \leq d \leq D(T)$. Moreover, it keeps the depth of the resulting sub-tree rooted in $n$ equal to that of the original sub-tree.
Figure~\ref{fig:splitting_reallocation_example} illustrates that procedure in which the splitting index of the (colored) node $n$ (on the left) is changed to originate the neighboring tree on the right. 

\subsection{Sub-region transfer neighborhood}
\label{sec:sub-region transfer}

\begin{figure}[h]
    \centering
	\includegraphics[width=\textwidth]{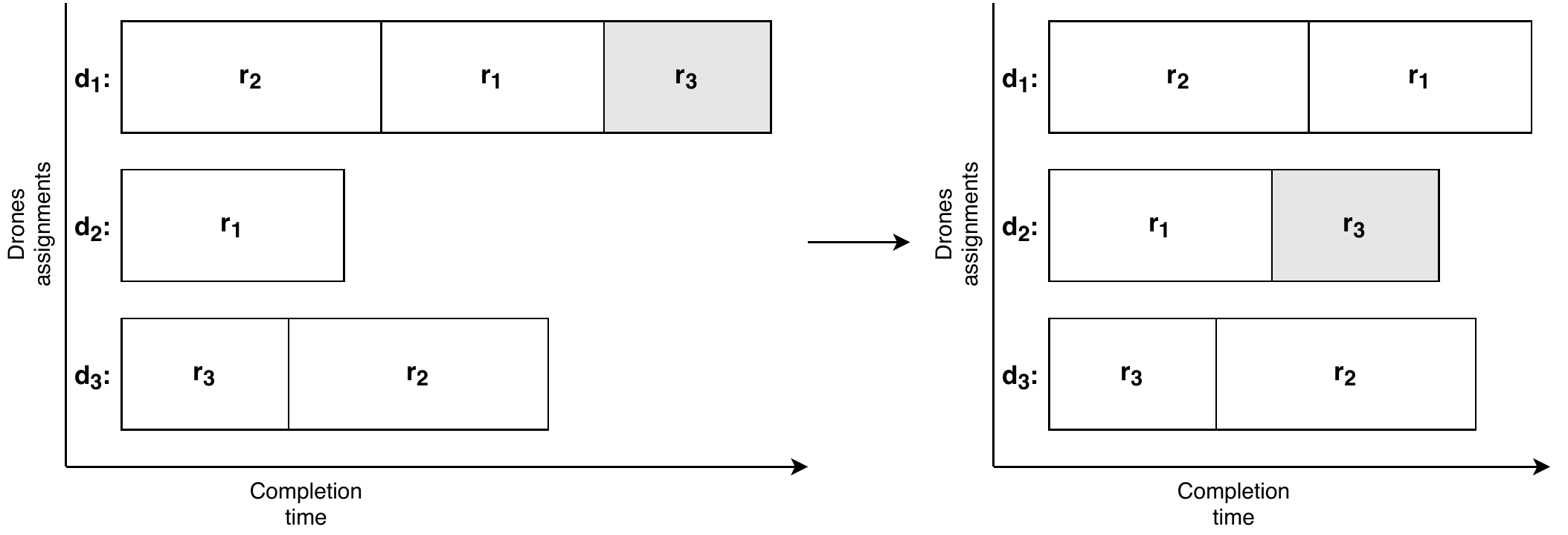}
   	\caption{Neighboring solution (right) obtained from a spatial partition tree (left) in the sub-region transfer neighborhood for $\sigma=2$.}
   	\label{fig:example_sub-region_transfer}
\end{figure}

This neighborhood encompasses all solutions obtained by transferring one sub-region assignment $(r_{\ell}, d_i)$ from a drone $d_i$ to a drone $d_j$ such that $i \neq j$.
In Figure~\ref{fig:example_sub-region_transfer}, the transfer of the sub-region $r_3$ from drone $d_1$ to drone $d_2$ generates a neighboring solution according to the sub-region transfer neighborhood for a reliability factor $\sigma$ equal to two.

\subsection{Sub-region swap neighborhood}
\label{sec:sub-region swap}

\begin{figure}[h]
    \centering
	\includegraphics[width=\textwidth]{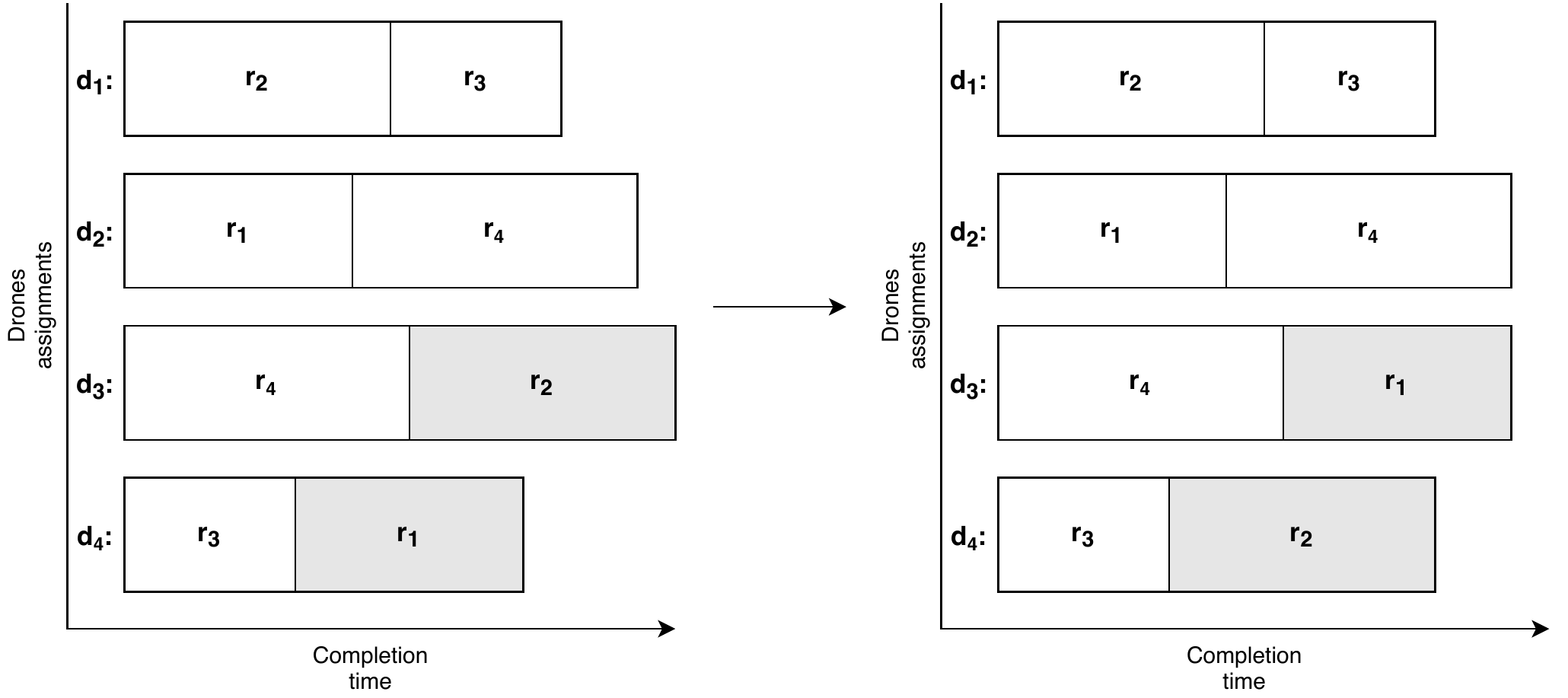}
   	\caption{Neighboring solution (right) obtained from a spatial partition tree (left) in the sub-region swap neighborhood for $\sigma=2$}
   	\label{fig:example_sub-region_swap}
\end{figure}

Given a solution, its neighbors in the sub-region swap neighborhood are obtained by the swap of sub-regions assignments $(r_{\ell}, d_i)$ and $(r_{m}, d_j)$ between  drones $d_i$ and $d_j$.
Such swap yields the sub-region assignments $(r_{\ell}, d_j)$ and $(r_{m}, d_i)$. 
An example of neighboring solution (for $\sigma=2$) in the sub-region swap neighborhood is given in Figure~\ref{fig:example_sub-region_swap}, where the assignments $(r_2, d_3)$ and $(r_1, d_4)$ are swapped between drones $d_3$ and $d_4$.

\subsection{Computation of the processing time and transmission data}
\label{sec:computation_times_and_data}

In order to efficiently explore the neighborhoods during VNS' local descent, it is important to adequately compute the photo processing time of a sub-region and the amount of data involved in the data transfers between drones.
Given the boundaries of a sub-region and the drone $d \in \bar{D}$ dealing with the reconstruction of that sub-region, this information can be computed in constant time as follows.

The computation of the photo processing time of a sub-region is based on the sets of photos $S^{left}_{c} $, $S^{right}_{c}$, $S^{up}_{\ell} $, $S^{down}_{\ell}$ for $c \in C$ and $\ell \in L$.
Thus,
\begin{align}
& S^{left}_{c} = \{ p \in P | lng(p) < lng(c) \}, \label{eq:s1} \\
& S^{right}_{c} = \{ p \in P | lng(p) > lng(c) \},\label{eq:s2} \\
& S^{up}_{\ell} = \{ p \in P | lat(p) < lat(\ell) \}, \label{eq:s3}\\ 
& S^{down}_{\ell} = \{ p \in P | lat(p) > lat(\ell) \} \label{eq:s4},
\end{align}
\noindent where $lat()$ and $lng()$ stand for the latitude and longitude respectively.
Thus, $S^{left}_{c} $ encompasses all photos on the left of $c$ whereas $S^{right}_{c}$ contains the photos on the right of $c$.
Similarly, $S^{up}_{\ell}$ groups the photos above $\ell$ and $S^{down}_{\ell} $ the photos below $\ell$.

We define $T(\overline{P})$ as the overall processing time of all photos in a set $\overline{P} \subseteq P$ 
such that
\begin{equation}
T(\overline{P}) = \sum_{p \in \overline{P}} \lambda_p.
\end{equation}
Finally, let $c_{<}$, $c_{>}$, $\ell_{\vee}$, $\ell_{\wedge}$ represent, respectively, the left, the right, the bottom (inferior), and the top (superior) boundaries of a sub-region $r$.
The photo processing time $T^{r}$ of a sub-region $r$ delimited by $c_{<}$, $c_{>}$, $\ell_{\vee}$, and $\ell_{\wedge}$  is then given by
\begin{equation}
T^{r} = T \left( \left\lbrace p \in P \mid 
lng(c_{<}) \leq lng(p) \leq lng(c_{>}) 
\text{ , } 
lat(\ell_{\vee}) \leq lat(p) \leq lat(\ell_{\wedge}) \right\rbrace \right). \nonumber
\end{equation}
Equivalently, one may subtract the processing times of the photos outside the sub-region $r$ from the overall processing time $T(P)$, i.e.,
\begin{equation}
T^{r} = T(P)  - T( S^{left}_{c_{<}} \cup S^{right}_{c_{>}} \cup S^{down}_{\ell_{\vee}}  \cup S^{up}_{\ell_{\wedge} } ).
\label{eq:photo_time_0}
\end{equation}

Now, for each pair $(c, \ell) \in C \times L$, let us define $Q^{1}_{c \ell}$, $Q^{2}_{c \ell}$, $Q^{3}_{c \ell}$, $Q^{4}_{c \ell}$, respectively, as the set of photos lying in the first, second, third, and fourth quadrants concerning $lng(c)$ and $lat(\ell)$, which are expressed as
\begin{align}
& Q^{1}_{c \ell} = \left\lbrace p \in P \mid lng(p) > lng(c), lat(p) > lat(\ell) \right\rbrace, \label{eq:q1} \\
& Q^{2}_{c \ell} = \left\lbrace p \in P \mid lng(p) < lng(c), lat(p) > lat(\ell) \right\rbrace, \label{eq:q2} \\
& Q^{3}_{c \ell} = \left\lbrace p \in P \mid lng(p) < lng(c), lat(p) < lat(\ell) \right\rbrace \text{, and} \label{eq:q3} \\
& Q^{4}_{c \ell} = \left\lbrace p \in P \mid lng(p) > lng(c), lat(p) < lat(\ell) \right\rbrace. \label{eq:q4}
\end{align}
Thus, it is possible to express equation (\ref{eq:photo_time_0}) as
\begin{equation}
\begin{split}
T^{r} = T(P)  - T( S^{left}_{c_{<}} ) - T(S^{right}_{c_{>}} ) - T(S^{down}_{\ell_{\vee}} ) - T(S^{up}_{\ell_{\wedge} } )
\\
+ T(Q^{1}_{ c_{>} \ell_{\wedge} }) + T(Q^{2}_{ c_{<} \ell_{\wedge} } ) + T(Q^{3}_{ c_{<} \ell_{\vee} } ) + T(Q^{4}_{ c_{>} \ell_{\vee} } ).
\end{split} 
\label{eq:photo_time}
\end{equation}

Given the boundaries of a sub-region $r$, equation (\ref{eq:photo_time}) allows to compute $T^{r}$ based only on the processing times of $S^{left}_{c_{<}}$, $S^{right}_{c_{>}}$, $S^{down}_{\ell_{\vee}}$, $S^{up}_{\ell_{\wedge} }$, 
$Q^{1}_{ c_{>} \ell_{\wedge} }$,  $Q^{2}_{ c_{<} \ell_{\wedge} }$, $Q^{3}_{ c_{<} \ell_{\vee} }$, $Q^{4}_{ c_{>} \ell_{\vee} }$, and $T(P)$. By precomputing these values in time $O(|C| \cdot |L| \cdot |P|)$ (equiv. $O(|P|^3)$, there is no need of keeping in memory the index of photos belonging to $r$.
Thus, the photo processing time of any sub-region $r$ can be computed by (\ref{eq:photo_time}) in constant time $O(1)$. 

Likewise, the amount of data $\mu^{hd}_{r}$ on a certain data transmission $f^{hd} \in F$ concerning only the sub-region $r$ can be computed by a closed form expression. 
Given the transmission demand $f^{hd}$ and a set of photos $\overline{P}$ , let $\mu^{hd}(\overline{P})$ be the amount of data in $\overline{P}$ exchanged from drone $h$ to the drone $d$, where $d$ is the photo-processing drone of $\overline{P}$, which is expressed as
\begin{equation}
\mu^{hd}(\overline{P}) = \sum_{p \in \overline{P}}  \theta_{hp} \mu_{p}.
\end{equation}
Since the boundaries of the sub-region $r$ and the transmission demand $f^{hd}$ are known, the $\mu^{hd}_{r}$ can be computed by the equation
\begin{align}
\mu^{hd}_{r} = & \quad \mu^{hd} \left( \left\lbrace p \in P \mid 
lng(c_{<}) \leq lng(p) \leq lng(c_{>}) 
\text{ , } 
lat(\ell_{\vee}) \leq lat(p) \leq lat(\ell_{\wedge}) \right\rbrace \right)  \nonumber \\
\mu^{hd}_{r} = & \quad \mu^{hd}(P)  - \mu^{hd}( S^{left}_{c_{<}} \cup S^{right}_{c_{>}} \cup S^{down}_{\ell_{\vee}}  \cup S^{up}_{\ell_{\wedge} } )  \nonumber \\
\mu^{hd}_{r} = & \quad \mu^{hd}(P)  - \mu^{hd}( S^{left}_{c_{<}} ) - \mu^{hd}(S^{right}_{c_{>}} ) - \mu^{hd}(S^{down}_{\ell_{\vee}} ) - \mu^{hd}(S^{up}_{\ell_{\wedge} } )   \nonumber \\
&+ \mu^{hd}(Q^{1}_{ c_{>} \ell_{\wedge} }) + \mu^{hd}(Q^{2}_{ c_{<} \ell_{\wedge} } ) + \mu^{hd}(Q^{3}_{ c_{<} \ell_{\vee} } ) + \mu^{hd}(Q^{4}_{ c_{>} \ell_{\vee} } ). \label{eq:data}
\end{align}
\noindent The terms within the equation (\ref{eq:data}) can all be precomputed in time $O(|D|^{2} \cdot |C| \cdot |L| \cdot |P| )$ operations (equiv. $|D|^{2} \cdot |P|^ 3$). 
Thus, the amount of data transmitted can be obtained in constant time given the boundaries of the sub-region $r$.

Note that both equations (\ref{eq:photo_time}) and (\ref{eq:data}) are suited to the way sub-regions are represented by a spatial partition tree. 
It only suffices to know the boundaries of a sub-region to query the terms in (\ref{eq:photo_time}) and (\ref{eq:data}).

\subsection{Proposed VNS heuristic}  
\label{sec:method}

According to Algorithm~\ref{alg:vns}, we need to define how to  create an initial solution as well as the diversification and intensification steps. 
These are described in the following.

\subsubsection*{Initial solution}

The initial solution is created by (i) randomly building a spatial-partition tree with $\bar{D}$ sub-regions (as defined in Section~\ref{sec:spatial_part_tree}); and by (ii) randomly assigning the obtained sub-regions to the set of 3D-capable drones $\bar{D}$ according to the reliability factor $\sigma$. 
Thus, each 3D-capable drone has at least one sub-region to process.

\subsubsection*{Diversification step}

The diversification step uses the \emph{sub-tree reconstruction} neighborhood. 
The VNS parameter $k$ establishes the depth from which a node will be selected in the spatial-partition tree $T$. 
More precisely, a node from depth $D(T) - k$  is randomly selected to be reconstructed.
Finally, we adopt a $k_{step}$ equal to one and $k_{max}$ equal to the $D(T)$ of the current best solution.

\subsubsection*{Intensification step}

Our local descent method is a Variable Neighborhood Descent (VND)~\cite{pierre_nenad2001vns}.
The proposed VND minimizes $T_{\max}$ by sequentially exploring the \textit{sub-region transfer}, \textit{sub-region swap}, and \textit{splitting hyperplane reallocation} neighborhoods.
In order to accelerate the local searches, we limit neighborhood exploration to transfer/swapping/reallocation of sub-regions assigned to the set of drones, named $D_{\max}$, holding the incumbent longest total processing time. 
That is, $D_{\max}$ comprises the drones whose total photo-processing time  is equal to the makespan ($T_{\max}$).
By denoting $R_{ \hat{D} }$ as the set of all sub-regions assigned to drones $d \in \hat{D} \subseteq D$, the local searches are implemented in such a way that
\begin{enumerate}
\item[i] $\mathcal{N}_{1}$ - \textbf{sub-region transfer}: switches sub-regions $r \in R_{D_{\max} }$ to other drones $d \in \bar{D} - D_{\max}$. 
The size of this neighborhood is bounded by $\Theta(|R_{ D_{\max} }| \times | \bar{D} - D_{max} |)$.

\item[ii] $\mathcal{N}_{2}$ - \textbf{sub-region swap}: only considers sub-region swaps ($r_{i}, r_{j}$) such that $r_{i} \in R_{D_{\max} }$.
Consequently, this neighborhood has $\Theta(|R_{ D_{\max} }| \times |R_{\bar{D} - D_{\max}}| )$ neighboring solutions.

\item[iii] $\mathcal{N}_{3}$ - \textbf{splitting hyperplane reallocation}: for each $n \in N_{D_{\max} }$, where $N_{D_{\max} }$ is composed by the parent nodes of leaf nodes representing the sub-regions $r \in R_{D_{\max} }$, this neighborhood reallocates the splitting hyperplanes (i.e., their positions and/or orientations)
of non-leaf nodes in the sub-tree rooted at $n$ (denoted $T_n$). 
Let us define $L_n$ and $C_n$ as the set of enclosed latitudes on axis $L$ and longitudes on axis $C$ at node $n$, respectively.
In the worst case, the size of this neighborhood is limited by $O\left( \sum_{ n \in N_{D_{\max}} } \left( |L_{n}|+|C_{n}| \right) ^{ |N_{T_n}| } \right)$, where $N_{T_n}$ is the set of non-leaf nodes in the sub-tree $T_n$. 

\end{enumerate}
\noindent Employing such sequence of neighborhoods grants higher priority to adjustments in the sub-region to drone assignments, which are encompassed in smaller neighborhoods that cause less changes to solution's structure.

Besides minimizing the latest processing time $T_{\max}$, the local searches seek solutions respecting the maximum allowed transmission time $\hat{T}$.
Thus, the transmission times $t^{hd}$ of all active transmission demands $f^{hd} \in F$ must be at most $\hat{T}$ (according to constraints (\ref{eq:scf_2})).
This means that we need to check whether
\begin{equation}
	t^{hd} \leq \hat{T} \qquad \forall f^{hd} \in F \label{eq:t_hat_feasibility}
\end{equation}
in order to define if a neighboring solution is feasible.
Further, the transmission times $t^{hd}$ are computed by the equation
\begin{equation}
	t^{hd} = \frac{ \mu^{hd} }{ \phi^{hd} } = \frac{ \sum_{r \in R_{d} }  \mu^{hd}_{r} }{ \phi^{hd} }. \label{eq:t_hd}
\end{equation}
As described in Section~\ref{sec:computation_times_and_data}, likewise the processing times $T^{r}$, it is possible to compute $\mu^{hd}_{r}$ in constant time.
However, computing $\phi^{hd}$ means solving an MMF rate allocation problem.
Due to the fixed tree topology, one can solve an MMF rate allocation problem in $O(|D|^{2} \times (|D|-1))$ employing the \textit{water filling} method~\cite{bertsekas1992data}.

Algorithm~\ref{alg:ls_first_improv} describes the local search used in this paper. 
It uses a first improvement exploration strategy, which pursues the first improvement direction in the local descent method. 
Limited computational experiments revealed that a best improving strategy in which the local descent used the best estimated improving direction was not better than its first improving counterpart.
The algorithm requires an initial solution $x$ and a neighborhood $\mathcal{N}_{t}$ as input and return the local optimum w.r.t. $\mathcal{N}_{t}$.
Also, let us define $\Phi$ as the hash table mapping a set of active transmission demands $\bar{F}$ to its respective MMF rate allocation $\phi$.
This data structure accelerates the local search since it prevents the method from solving the same MMF rate allocation problem more than once for the same set of active transmission demands.
Finally, let us define the function $T_{\max}(\cdot)$ to return the makespan (i.e., latest completion time) of a solution in its argument.

\begin{algorithm}[h]
\caption{Local search using first improvement search strategy.}
\label{alg:ls_first_improv}
\begin{algorithmic}[1]
\medskip
\Require $x$ and $\mathcal{N}_{t}$
\State $Stop \leftarrow false$;
	\While{not $Stop$}
		\State $Stop \leftarrow True$;
		\ForAll{$x^{\prime} \in \mathcal{N}_{t}(x) $}
			\If{$T_{\max}(x^{\prime}) < T_{\max}(x)$}
				\State identify the active demands $\bar{F}(x^{\prime})$;
				\If{ $\nexists$ $\Phi( \bar{F}(x^{\prime}) )$}
					\State $\Phi( \bar{F}(x^{\prime}) ) \leftarrow$ \emph{water filling algorithm}$(\bar{F}(x^{\prime})$;
				\EndIf
				\State $\bar{T} \leftarrow$ compute communication delays $(\Phi( \bar{F}(x^{\prime}) ), \mu(x^{\prime}) )$;	
				\If{$t^{hd} \leq \hat{T} \quad \forall t^{hd} \in \bar{T} $}
					\State $Stop \leftarrow False$;
					\State $x \leftarrow x^{\prime}$;
					\State \textbf{break};
				\EndIf
			\EndIf
		\EndFor
	\EndWhile
\State
\Return $x$;
\end{algorithmic}
\end{algorithm}

Regarding Algorithm~\ref{alg:ls_first_improv}, step~1 initializes the flag $Stop$ to false. 
While the local optimum is not reached (steps~2-18), the local search explores the neighboring solutions $x^{\prime} \in \mathcal{N}_{t}(x)$ (steps~4-17).
If for neighboring solution $x^{\prime}$ $T_{\max}(x^{\prime}) < T_{\max}(x)$ (steps~5-16), the algorithm proceeds to checking whether $x^{\prime}$ has feasible transmission times associated with it according to expression~(\ref{eq:t_hat_feasibility}).
First, it identifies the active demands $\bar{F}$ in $x^{\prime}$ (step~6).
A demand between a pair of drones is active if there exists any data to be exchanged between them~(Section \ref{sec:math_formulation}).
Thus, step~6 checks whether $\mu^{hd} = \sum_{r \in R_{d}}  \mu^{hd}_{r} > 0$ for each $f^{hd} \in F$.
Then, it computes the MMF rate allocation problem for $\bar{F}$ (through the \emph{water filling algorithm}) if and only if $\bar{F}$ is not already a stored key of $\Phi$ (steps~7-9).
The set of communication delays $\bar{T} = \{t^{hd} | f^{hd} \in F\}$  is computed in step~10 according to the equation (\ref{eq:t_hd}) given the transmission rates $\Phi( \bar{F}(x^{\prime}))$ and the amount of  data $\mu = \{ \mu^{hd} | f^{hd} \in F \}$ in  solution $x^{\prime}$.
If $x^{\prime}$ has all communication delays $t^{hd} \in \bar{T}$ smaller or equal to $\hat{T}$ (steps~11-15), the local search
(i) marks $x$ as sub-optimum (step~12); 
(ii) sets $x^{\prime}$ as the new $x$ (step~13), and 
(iii) breaks the for loop to explore the neighborhood w.r.t. the new incumbent $x$ (step~14). 
Once reached, the local optimum concerning $\mathcal{N}_{t}$ is returned in step~19.

Finally, the proposed VND is summarized in Algorithm~\ref{alg:vnd}.
\begin{algorithm}[h]
\caption{Variable Neighborhood Descend(VND) for the CAPsac}
\label{alg:vnd}
\begin{algorithmic}[1]
\medskip
\State $\mathcal{N} = \{$\textit{sub-region transfer}, \textit{sub-region swap}, \textit{splitting hyperplane reallocation} $\}$
\State $t \leftarrow 1$
\If{$\sigma =1$}
 \State $t \leftarrow 2$
\EndIf
\Repeat
\State $x^{\prime \prime} \leftarrow$ perform LocalSearch($x^{\prime}$, $\mathcal{N}_{t}$);
\If{$T_{\max}(x^{\prime\prime}) < T_{\max}(x^{\prime})$}
  \State $x^{\prime} \leftarrow x^{\prime\prime}$;
  \State $t \leftarrow 1$
  	\If{$\sigma =1$}
 		\State $t \leftarrow 2$
	\EndIf
  \Else
  \State $t \leftarrow t + 1$;
\EndIf
\Until{$t > 3$}
\end{algorithmic}
\end{algorithm}
\noindent 
In the steps~2-5 of the algorithm, we choose the initial value of $t$ according to the reliability factor ($\sigma$) used. 
It is set to $t=1$ for $\sigma > 1$ and $t=2$ otherwise. 
That is, the local search applying the \textit{sub-region transfer} neighborhood ($\mathcal{N}_{1}$) is used only for instances with $\sigma > 1$ since it cannot find better solutions for $\sigma =1$. 
This is due to the fact that each 3D-capable drone must have at least one sub-region assigned to it.  
The VND proceeds by applying the local searches according to the current neighborhood $\mathcal{N}_{t}$ until it reaches the local optimum w.r.t. all neighborhoods in $\mathcal{N}$ (steps~6-17).
Each new solution $x^{\prime \prime}$ (obtained at step~7 by Algorithm~\ref{alg:ls_first_improv}) with better $T_{\max}$ than $x^{\prime}$ is kept and $t$ is reset likewise steps~2-5 (steps~8-13).
Otherwise, $t$ is increased by one (steps~14-16), and the algorithm iterates.

\section{Computational experiments}
\label{sec:experiments}

Our experimental analysis aims to evaluate 
(i) the effectiveness of the decomposition strategy;
(ii) the effectiveness of employing different neighborhoods in our variable neighborhood method; and
(iii) the performance of the proposed methods and sensitivity regarding both the reliability factor $\sigma$ and the maximum transmission time allowed $\hat{T}$.
We used CPLEX~v12.10 as general-purpose integer linear programming solver and C++ as programming language compiled with gcc~v5.4.0.  
All experiments were carried exploiting a single core on a machine powered by an Intel E5-2683~v4 Broadwell 2.1GHz with 20Gb of RAM, and running the CentOS Linux 7.5.1804 OS.

\subsection{Instances and notation adopted}

Our experimental analysis is carried on the instances proposed in~\citet{capsac2020} (available at \url{https://github.com/ds4dm/CAPsac}). 
All instances are based on realistic data and comprise two scenarios: 
(i) one in which all photos are processed in the same amount of time $\lambda$, named \textit{unweighted}; 
and (ii) another one in which each photo has a different processing time $\lambda_p$, named \textit{weighted}. 
The name of the instances follows the notation~\textit{X}-P\textit{Y}D\textit{Z}$\%\bar{D}$\textit{W} where “X” is “u” for the unweighted instances and “w” for the weighted instances, “Y” stands for the number of photos in the instance, “Z” specifies the number of drones in the swarm, and “W” informs the percentage of drones able to perform 3D reconstruction. 
Remark that the number of 3D-capable drones ($|\bar{D}|$) is always equal to $\left\lfloor  Z \times \frac{W}{100} \right\rfloor $. 
The characteristics of the tested instances are listed in Table \ref{tab:instances}.

\begin{table}
\centering
\caption{Characteristics of the tested instances.}
\label{tab:instances}
\begin{tabular}{l | l}
	\hline\noalign{\smallskip}
  	Images(P) &  200, 400, 500, 750, 1000 \\
  	Drones(D) & 5, 7, 10, 15 \\ 
  	$\%$3D-capable drones($\%\bar{D}$) & 50\%, 70\%, 90\% \\
    \noalign{\smallskip}\hline
\end{tabular}
\end{table}

Optimal makespans are obtained by solving the CAPsac formulation~\cite{capsac2020} with CPLEX within one day of execution.
Whenever CPLEX is not able to prove optimality within this time horizon, the best feasible solution obtained in~\cite{capsac2020} is reported instead, which is indicated by an “$\ast$” in the tables. 
Average and best results are then computed according to that best known solution.

\subsection{Effectiveness of the decomposition strategy}
\label{sec:decomp_vs_rcapsac}

This section assesses the effectiveness of the decomposition strategy exploited by the proposed decomposition-based heuristic. 

Tables~\ref{tab:rcapsac_x_decomp_unweighted} and \ref{tab:rcapsac_x_decomp_weighted} report for each instance (named under column ``Instance'') the following values regarding the optimal (or best known) solutions:
\begin{itemize}
    \item[-] the smallest cardinality found among the sub-regions (column $n^{*}_{\ell}$);
    \item[-] the largest cardinality found among the sub-regions (column $n^{*}_{u}$); 
    \item[-] the cardinality of the sub-region which yields the makespan (column $n^{*}_{mks}$);
    \item[-] the associated makespan value (column \textsc{opt}).
\end{itemize}

Further, we report the percentage deviations (columns ``Dev.(\%)'') achieved by the complete $rCAPsac$ formulation and the decomposition method with respect to the best known makespan. 
We also report the total computing CPU times spent by both methods under column (``sec''). 
Finally, concerning the decomposition-based heuristic, we report the last interval $[n_{\ell}, n_{u}  ]$ used before the algorithm is halted by its stopping condition.
For all experiments,  we set a time limit of 2 hours for the MIP formulation solved by CPLEX in each iteration of the decomposition-based heuristic, and 24 hours as maximum time for solving rCAPsac by CPLEX.
A trace (``$-$'') is reported whenever CPLEX is not able to find a feasible solution within this time limit.

\begin{table}[H]
\centering
\setlength{\tabcolsep}{2pt}
\footnotesize
\caption{Percentage deviations of the complete $rCAPsac$ formulation and the decomposition method when solving unweighted instances.}
\label{tab:rcapsac_x_decomp_unweighted}
\begin{tabular}{ l c r r r | r  r | r r r r }
	\hline\noalign{\smallskip}
	\multicolumn{5}{l}{  $\sigma=1, \hat{T}=\infty$, stop condition$=86400s$ } 
	& \multicolumn{2}{  c }{ $rCAPsac$}
	& \multicolumn{4}{  c }{ Decomposition}	\\
   \noalign{\smallskip}\hline\noalign{\smallskip}
	Instance & $n^{*}_{\ell}$ & $n^{*}_{u}$ & $n^{*}_{mks}$ &  \multicolumn{1}{ r}{\textsc{opt}}
	& \multicolumn{1}{ r}{Dev.(\%)} & \multicolumn{1}{r}{sec.} 
	& \multicolumn{1}{ r}{$n_{\ell}$} & $n_{u}$ & \multicolumn{1}{ r}{Dev.(\%)}  & \multicolumn{1}{r}{sec.}  \\
   \noalign{\smallskip}\hline\noalign{\smallskip} 
	u-P200D5$\%\bar{D}$70 & 65 & 70 & 70 & 1870.40 & 0.00 & 179.37 & 64 & 72 & 0.00 & 0.28 \\ 
u-P200D7$\%\bar{D}$50 & 60 & 70 & 70 & 1870.40 & 0.00 & 548.55 & 64 & 72 & 0.00 & 0.50 \\ 
u-P400D5$\%\bar{D}$70 & 132 & 135 & 135 & 3607.20 & 0.00 & 6996.32 & 128 & 140 & 0.00 & 1.04 \\ 
u-P400D7$\%\bar{D}$50 & 130 & 135 & 135 & 3607.20 & 0.00 & 11355.80 & 128 & 140 & 0.00 & 1.55 \\ 
\hline 
u-P200D5$\%\bar{D}$90 & 50 & 50 & 50 & 1336.00 & 0.00 & 244.25 & 49 & 51 & 0.00 & 0.11 \\ 
u-P200D7$\%\bar{D}$70 & 50 & 50 & 50 & 1336.00 & 0.00 & 256.03 & 49 & 51 & 0.00 & 0.23 \\ 
u-P400D5$\%\bar{D}$90 & 100 & 100 & 100 & 2672.00 & 0.00 & 9346.54 & 99 & 102 & 0.00 & 0.46 \\ 
u-P400D7$\%\bar{D}$70 & 100 & 100 & 100 & 2672.00 & 0.00 & 45059.71 & 99 & 102 & 0.00 & 0.55 \\ 
\hline 
u-P200D10$\%\bar{D}$50 & 40 & 40 & 40 & 1068.80 & 0.00 & 1838.53 & 39 & 42 & 0.00 & 1.46 \\ 
u-P400D10$\%\bar{D}$50 & 80 & 80 & 80 & 2137.60 & 0.00 & 43171.40 & 78 & 81 & 0.00 & 5.44 \\ 
\hline 
u-P200D7$\%\bar{D}$90 & 30 & 34 & 34 & 908.48 & 0.00 & 1901.86 & 28 & 38 & 0.00 & 8.00 \\ 
u-P400D7$\%\bar{D}$90 & 60 & 68 & 68 & *1816.96 & 32.35 & 86400.00 & 64 & 70 & 1.47 & 13.28 \\ 
\hline 
u-P200D10$\%\bar{D}$70 & 24 & 30 & 30 & *801.60 & 33.33 & 86400.00 & 27 & 32 & 0.00 & 115.18 \\ 
u-P400D10$\%\bar{D}$70 & 50 & 60 & 60 & *1603.20 & 300.00 & 86400.00 & 54 & 63 & -3.33 & 2013.98 \\ 
u-P500D15$\%\bar{D}$50 & 70 & 72 & 72 & 1923.84 & - & 86400.00 & 69 & 75 & 0.00 & 333.59 \\ 
u-P750D15$\%\bar{D}$50 & 108 & 117 & 117 & *3126.24 & - & 86400.00 & 104 & 110 & -7.69 & 262.73 \\ 
u-P1000D15$\%\bar{D}$50 & 143 & 144 & 144 & *3847.68 & - & 86400.00 & 136 & 145 & 0.00 & 7434.67 \\ 
\hline 
u-P200D10$\%\bar{D}$90 & 18 & 24 & 24 & *641.28 & 8.33 & 86400.00 & 21 & 25 & 0.00 & 3366.98 \\ 
u-P400D10$\%\bar{D}$90 & 42 & 45 & 45 & 1202.40 & - & 86400.00 & 42 & 46 & 0.00 & 3262.06 \\ 
\noalign{\smallskip}\hline

\end{tabular}
\end{table}

\begin{table}[H]
\centering
\setlength{\tabcolsep}{2pt}
\footnotesize
\caption{Percentage deviations of the complete $rCAPsac$ formulation and the decomposition method when solving weighted instances.}
\label{tab:rcapsac_x_decomp_weighted}
\begin{tabular}{ l c r r r | r  r | r r r r }
	\hline\noalign{\smallskip}
	\multicolumn{5}{l}{  $\sigma=1, \hat{T}=\infty$, stop condition$=86400s$ } 
	& \multicolumn{2}{  c }{ $rCAPsac$}
	& \multicolumn{4}{  c }{ Decomposition}	\\
   \noalign{\smallskip}\hline\noalign{\smallskip}
	Instance & $n^{*}_{\ell}$ & $n^{*}_{u}$ & $n^{*}_{mks}$ &  \multicolumn{1}{ r}{\textsc{opt}}
	& \multicolumn{1}{ r}{Dev.(\%)} & \multicolumn{1}{r}{sec.} 
	& \multicolumn{1}{ r}{$n_{\ell}$} & $n_{u}$ & \multicolumn{1}{ r}{Dev.(\%)}  & \multicolumn{1}{r}{sec.}  \\
   \noalign{\smallskip}\hline\noalign{\smallskip}
	w-P200D5$\%\bar{D}$70 & 65 & 70 & 70 & 1886.98 & 0.00 & 242.54 & 64 & 72 & 0.00 & 0.36 \\ 
w-P200D7$\%\bar{D}$50 & 60 & 70 & 70 & 1885.31 & 0.00 & 784.93 & 64 & 72 & 3.26 & 0.39 \\ 
w-P400D5$\%\bar{D}$70 & 130 & 135 & 135 & 3735.32 & 0.00 & 10209.88 & 128 & 140 & 0.00 & 1.21 \\ 
w-P400D7$\%\bar{D}$50 & 130 & 135 & 135 & 3773.12 & 0.00 & 5987.39 & 128 & 140 & 0.00 & 1.93 \\ 
\hline 
w-P200D5$\%\bar{D}$90 & 50 & 50 & 50 & 1409.48 & 0.00 & 212.44 & 49 & 51 & 0.00 & 0.30 \\ 
w-P200D7$\%\bar{D}$70 & 50 & 50 & 50 & 1399.82 & 0.00 & 802.39 & 49 & 51 & 0.00 & 0.56 \\ 
w-P400D5$\%\bar{D}$90 & 100 & 100 & 100 & 2812.93 & 0.00 & 21189.45 & 99 & 102 & 0.00 & 0.57 \\ 
w-P400D7$\%\bar{D}$70 & 100 & 100 & 100 & 2791.66 & 0.00 & 40872.49 & 99 & 102 & 0.00 & 1.49 \\ 
\hline 
w-P200D10$\%\bar{D}$50 & 40 & 40 & 40 & 1121.89 & 0.00 & 9106.51 & 39 & 42 & 0.00 & 13.92 \\ 
w-P400D10$\%\bar{D}$50 & 72 & 88 & 88 & 2290.48 & 0.00 & 40443.02 & 78 & 81 & 1.15 & 34.01 \\ 
\hline 
w-P200D7$\%\bar{D}$90 & 30 & 35 & 35 & 944.06 & 0.00 & 8669.90 & 28 & 38 & 0.00 & 17.77 \\ 
w-P400D7$\%\bar{D}$90 & 60 & 75 & 75 & *1909.18 & 19.17 & 86400.00 & 63 & 72 & -0.94 & 25.17 \\ 
\hline 
w-P200D10$\%\bar{D}$70 & 26 & 30 & 30 & 820.40 & 0.00 & 62142.01 & 27 & 32 & 0.29 & 1086.76 \\ 
w-P400D10$\%\bar{D}$70 & 56 & 63 & 60 & *1671.12 & 123.18 & 86400.00 & 55 & 62 & -2.30 & 7232.05 \\ 
w-P500D15$\%\bar{D}$50 & 66 & 78 & 72 & *2028.11 & - & 86400.00 & 69 & 75 & -2.40 & 1791.89 \\ 
w-P750D15$\%\bar{D}$50 & 96 & 119 & 117 & *3138.73 & - & 86400.00 & 102 & 112 & -4.91 & 2562.77 \\ 
w-P1000D15$\%\bar{D}$50 & 132 & 160 & 160 & *4343.26 & - & 86400.00 & 135 & 147 & -9.80 & 7762.67 \\ 
\hline 
w-P200D10$\%\bar{D}$90 & 18 & 24 & 24 & *665.54 & 20.03 & 86400.00 & 19 & 27 & -2.85 & 15655.03 \\ 
w-P400D10$\%\bar{D}$90 & 40 & 50 & 50 & *1302.82 & - & 86400.00 & 39 & 49 & -2.89 & 9645.66 \\ 
\noalign{\smallskip}\hline

\end{tabular}
\end{table}

The decomposition strategy proves effective and yields equivalent or better deviation values than solving the complete $rCAPsac$ formulation by CPLEX within the defined time limit. 
In fact, it is able to find a best known makespan for several instances, as indicated by the negative deviations in the table.
However, for a couple of instances, i.e., (“$u{-}P400D7\%D90$”,  “$w{-}P200D7\%D50$”,  “$w{-}P400D10\%D50$”,  “$w{-}P200D10\%D70$”), the decomposition algorithm is not able to attain the best known solutions obtained in~\cite{capsac2020}.  
For those cases, we observe that the intervals 
$[n_{\ell}, n_{u}  ]$ must be increased to allow subsets of photos whose cardinalities are out of the defined bounds. For example, for instance $u{-}P400D7\%D90$, the best known solution contains a sub-region with 60 photos while the last iteration of the decomposition method was executed for $[n_\ell,n_u] = [64,70]$.   Finally, the decomposition method demands significant smaller execution times than solving the complete $rCAPsac$ formulation by CPLEX as well as, in general, the formulations in~\cite{capsac2020}.

\vspace{-3mm}
\subsection{Effectiveness of the neighborhoods}
\label{sec:neighborhoods_analysis}

To assess the sequence of neighborhoods in our proposed VND, we compare the performance of distinct neighborhood configurations ---  $\mathcal{N}_{1}$, $\mathcal{N}_{1}+\mathcal{N}_{2}$, $\mathcal{N}_{1}+\mathcal{N}_{3}$, $\mathcal{N}_{1}+\mathcal{N}_{2}+\mathcal{N}_{3}$ --- within our proposed VNS method, where
$\mathcal{N}_{1}$ refers to the \emph{sub-region transfer} neighborhood, $\mathcal{N}_{2}$ refers to \emph{sub-region swap} neighborhood, and $\mathcal{N}_{3}$ refers to the \emph{splitting hyperplane reallocation} neighborhood.

Tables~\ref{tab:neighborhoods_unweighted} and \ref{tab:neighborhoods_weighted} report the results of the unweighted and weighted instances, respectively. 
For each ``Instance'' and reliability factor ``$\sigma$'', the column ``\textsc{min.}'' presents the shortest makespan found across all neighborhood configurations.
The average percentage deviation from ``\textsc{min.}'' achieved by each neighborhood configuration is reported in columns ``Avg$(\%)$''.

The results are computed for 20 executions of the VNS-based heuristics within 300s of CPU execution time.  
These experiments are conducted over different reliability factors $\sigma > 1$ so as that the \emph{sub-region transfer} neighborhood could be evaluated.
Moreover, we do not present  results for instances with fewer than five 3D-capable drones since all neighborhood configurations performed equivalently.

\begin{table}[H]
\centering
\footnotesize
\setlength{\tabcolsep}{1.3pt}
\caption{Percentage deviations of employing distinct neighborhoods in the VND when solving unweighted instances.}
\label{tab:neighborhoods_unweighted}
\begin{tabular}{ l c r | r | r | r | r }
	\hline\noalign{\smallskip}
	\multicolumn{3}{l}{  \multirow{2}{*}{$\hat{T}=\infty$, stop condition$=300s$}  } & \multicolumn{4}{c}{VNS}  \\
	\noalign{\smallskip}\cline{4-7}\noalign{\smallskip}
	& & \multicolumn{1}{c}{} & \multicolumn{1}{c}{$\mathcal{N}_{1}$} 
	& \multicolumn{1}{c}{$\mathcal{N}_{1}+\mathcal{N}_{2}$}
	& \multicolumn{1}{c}{$\mathcal{N}_{1}+\mathcal{N}_{3}$}
	& \multicolumn{1}{c}{$\mathcal{N}_{1}{+}\mathcal{N}_{2}{+}\mathcal{N}_{3}$} \\
   \noalign{\smallskip}\hline\noalign{\smallskip}
	Instance & $\sigma$ & \multicolumn{1}{r}{\textsc{min.}}
	& \multicolumn{1}{r}{Avg.(\%)}
	& \multicolumn{1}{r}{Avg.(\%)}
	& \multicolumn{1}{r}{Avg.(\%)}
	& \multicolumn{1}{r}{Avg.(\%)} \\
   \noalign{\smallskip}\hline\noalign{\smallskip}
	 \multirow{3}{*}{u-P200D10$\%\bar{D}$50} & 2 & 2137.60 & \textbf{0.00} & \textbf{0.00} & \textbf{0.00} & \textbf{0.00} \\ 
 & 3 & 3206.40 & \textbf{0.00} & \textbf{0.00} & \textbf{0.00} & \textbf{0.00} \\ 
 & 4 & 4275.20 & \textbf{0.00} & \textbf{0.00} & \textbf{0.00} & \textbf{0.00} \\ 
\hline 
 \multirow{3}{*}{u-P400D10$\%\bar{D}$50} & 2 & 4275.20 & \textbf{0.00} & \textbf{0.00} & \textbf{0.00} & \textbf{0.00} \\ 
 & 3 & 6412.80 & \textbf{0.00} & \textbf{0.00} & \textbf{0.00} & \textbf{0.00} \\ 
 & 4 & 8550.40 & \textbf{0.00} & \textbf{0.00} & \textbf{0.00} & \textbf{0.00} \\ 
\hline 
\hline 
 \multirow{3}{*}{u-P200D7$\%\bar{D}$90} & 2 & 1790.24 & \textbf{0.00} & \textbf{0.00} & \textbf{0.00} & \textbf{0.00} \\ 
 & 3 & 2672.00 & \textbf{0.00} & \textbf{0.00} & \textbf{0.00} & \textbf{0.00} \\ 
 & 4 & 3580.48 & \textbf{0.00} & \textbf{0.00} & \textbf{0.00} & \textbf{0.00} \\ 
\hline 
 \multirow{3}{*}{u-P400D7$\%\bar{D}$90} & 2 & 3580.48 & \textbf{0.00} & \textbf{0.00} & \textbf{0.00} & \textbf{0.00} \\ 
 & 3 & 5344.00 & \textbf{0.00} & \textbf{0.00} & \textbf{0.00} & \textbf{0.00} \\ 
 & 4 & 7134.24 & \textbf{0.00} & \textbf{0.00} & \textbf{0.00} & \textbf{0.00} \\ 
\hline 
\hline 
 \multirow{3}{*}{u-P200D10$\%\bar{D}$70} & 2 & 1549.76 & 0.86 & 0.09 & \textbf{0.00} & \textbf{0.00} \\ 
 & 3 & 2297.92 & 0.64 & 0.35 & 0.35 & \textbf{0.06} \\ 
 & 4 & 3072.80 & \textbf{0.04} & 0.17 & 0.09 & \textbf{0.04} \\ 
\hline 
 \multirow{3}{*}{u-P400D10$\%\bar{D}$70} & 2 & 3072.80 & 0.43 & 0.52 & 0.04 & \textbf{0.00} \\ 
 & 3 & 4595.84 & 0.52 & 1.98 & 0.23 & \textbf{0.12} \\ 
 & 4 & 6118.88 & 0.39 & 0.44 & 0.39 & \textbf{0.31} \\ 
\hline 
 \multirow{3}{*}{u-P500D15$\%\bar{D}$50} & 2 & 3847.68 & 0.42 & 0.62 & \textbf{0.00} & \textbf{0.00} \\ 
 & 3 & 5744.80 & 0.56 & 1.23 & 0.23 & \textbf{0.05} \\ 
 & 4 & 7641.92 & 0.89 & 0.77 & 0.70 & \textbf{0.61} \\ 
\hline 
 \multirow{3}{*}{u-P750D15$\%\bar{D}$50} & 2 & 5744.80 & 1.02 & 1.26 & \textbf{0.44} & 0.47 \\ 
 & 3 & 8630.56 & 1.24 & 0.46 & 0.26 & \textbf{0.20} \\ 
 & 4 & 11489.60 & 0.64 & 0.42 & 0.23 & \textbf{0.16} \\ 
\hline 
 \multirow{3}{*}{u-P1000D15$\%\bar{D}$50} & 2 & 7641.92 & 1.17 & 1.01 & 0.65 & \textbf{0.52} \\ 
 & 3 & 11462.88 & 1.42 & 0.65 & \textbf{0.37} & 1.25 \\ 
 & 4 & 15283.84 & 0.62 & 0.64 & 0.52 & \textbf{0.43} \\ 
\hline 
\hline 
 \multirow{3}{*}{u-P200D10$\%\bar{D}$90} & 2 & 1202.40 & 0.56 & 0.11 & \textbf{0.00} & 0.11 \\ 
 & 3 & 1790.24 & \textbf{0.00} & \textbf{0.00} & \textbf{0.00} & \textbf{0.00} \\ 
 & 4 & 2378.08 & 1.12 & 1.01 & 1.12 & \textbf{0.96} \\ 
\hline 
 \multirow{3}{*}{u-P400D10$\%\bar{D}$90} & 2 & 2404.80 & 0.22 & 0.11 & \textbf{0.00} & \textbf{0.00} \\ 
 & 3 & 3580.48 & \textbf{0.00} & \textbf{0.00} & \textbf{0.00} & \textbf{0.00} \\ 
 & 4 & 4756.16 & 0.79 & 0.56 & 0.53 & \textbf{0.45} \\ 
\hline 
\hline 
\multicolumn{3}{ r|}{Average} & 0.41 & 0.38 & 0.19 & \textbf{0.17} \\ 
\noalign{\smallskip}\hline

\end{tabular}
\end{table}

\begin{table}[H]
\centering
\footnotesize
\setlength{\tabcolsep}{1.3pt}
\caption{Percentage deviations of employing distinct neighborhoods in the VND when solving weighted instances.}
\label{tab:neighborhoods_weighted}
\begin{tabular}{ l c r | r | r | r | r }
	\hline\noalign{\smallskip}
	\multicolumn{3}{l}{  \multirow{2}{*}{$\hat{T}=\infty$, stop condition$=300s$}  } & \multicolumn{4}{c}{VNS}  \\
	\noalign{\smallskip}\cline{4-7}\noalign{\smallskip}
	& & \multicolumn{1}{c}{} & \multicolumn{1}{c}{$\mathcal{N}_{1}$} 
	& \multicolumn{1}{c}{$\mathcal{N}_{1}+\mathcal{N}_{2}$}
	& \multicolumn{1}{c}{$\mathcal{N}_{1}+\mathcal{N}_{3}$}
	& \multicolumn{1}{c}{$\mathcal{N}_{1}{+}\mathcal{N}_{2}{+}\mathcal{N}_{3}$} \\
   \noalign{\smallskip}\hline\noalign{\smallskip}
	Instance & $\sigma$ & \multicolumn{1}{r}{\textsc{min.}}
	& \multicolumn{1}{r}{Avg.(\%)}
	& \multicolumn{1}{r}{Avg.(\%)}
	& \multicolumn{1}{r}{Avg.(\%)}
	& \multicolumn{1}{r}{Avg.(\%)} \\
   \noalign{\smallskip}\hline\noalign{\smallskip}
	 \multirow{3}{*}{w-P200D10$\%\bar{D}$50} & 2 & 2212.64 & 0.18 & 0.16 & 0.11 & \textbf{0.06} \\ 
 & 3 & 3314.91 & 0.16 & 0.18 & \textbf{0.15} & 0.16 \\ 
 & 4 & 4436.74 & 0.03 & 0.02 & \textbf{0.02} & \textbf{0.02} \\ 
\hline 
 \multirow{3}{*}{w-P400D10$\%\bar{D}$50} & 2 & 4433.53 & 0.19 & 0.22 & \textbf{0.08} & 0.09 \\ 
 & 3 & 6648.11 & 0.22 & 0.24 & \textbf{0.13} & 0.15 \\ 
 & 4 & 8901.19 & \textbf{0.01} & 0.03 & 0.03 & 0.02 \\ 
\hline 
\hline 
 \multirow{3}{*}{w-P200D7$\%\bar{D}$90} & 2 & 1816.30 & 0.14 & 0.16 & 0.04 & \textbf{0.03} \\ 
 & 3 & 2724.19 & 0.00 & 0.00 & \textbf{0.00} & \textbf{0.00} \\ 
 & 4 & 3632.45 & 0.06 & 0.06 & 0.03 & \textbf{0.02} \\ 
\hline 
 \multirow{3}{*}{w-P400D7$\%\bar{D}$90} & 2 & 3697.18 & 0.11 & 0.11 & \textbf{0.02} & \textbf{0.02} \\ 
 & 3 & 5545.48 & \textbf{0.00} & \textbf{0.00} & \textbf{0.00} & \textbf{0.00} \\ 
 & 4 & 7394.18 & 0.06 & 0.06 & \textbf{0.03} & \textbf{0.03} \\ 
\hline 
\hline 
 \multirow{3}{*}{w-P200D10$\%\bar{D}$70} & 2 & 1577.87 & 0.97 & 0.84 & 0.40 & \textbf{0.30} \\ 
 & 3 & 2365.78 & 0.63 & 0.49 & 0.43 & \textbf{0.26} \\ 
 & 4 & 3155.85 & 0.66 & 0.40 & 0.34 & \textbf{0.33} \\ 
\hline 
 \multirow{3}{*}{w-P400D10$\%\bar{D}$70} & 2 & 3164.05 & 0.65 & 0.91 & 0.47 & \textbf{0.33} \\ 
 & 3 & 4747.60 & 0.52 & 0.35 & 0.37 & \textbf{0.18} \\ 
 & 4 & 6326.53 & 0.57 & 0.40 & 0.35 & \textbf{0.32} \\ 
\hline 
 \multirow{3}{*}{w-P500D15$\%\bar{D}$50} & 2 & 3902.67 & 0.79 & 0.68 & \textbf{0.27} & \textbf{0.27} \\ 
 & 3 & 5852.50 & 1.33 & 1.18 & 0.27 & \textbf{0.21} \\ 
 & 4 & 7801.65 & 0.72 & 0.54 & 0.31 & \textbf{0.22} \\ 
\hline 
 \multirow{3}{*}{w-P750D15$\%\bar{D}$50} & 2 & 5843.52 & 0.67 & 0.99 & 0.28 & \textbf{0.27} \\ 
 & 3 & 8762.21 & 0.58 & 0.44 & 0.32 & \textbf{0.24} \\ 
 & 4 & 11679.59 & 0.70 & 0.55 & 0.32 & \textbf{0.28} \\ 
\hline 
 \multirow{3}{*}{w-P1000D15$\%\bar{D}$50} & 2 & 7724.25 & 1.08 & 1.04 & \textbf{0.34} & 0.37 \\ 
 & 3 & 11590.54 & 0.58 & 1.11 & 0.32 & \textbf{0.21} \\ 
 & 4 & 15451.40 & 0.58 & 0.36 & \textbf{0.19} & 0.26 \\ 
\hline 
\hline 
 \multirow{3}{*}{w-P200D10$\%\bar{D}$90} & 2 & 1230.42 & 1.10 & 1.08 & \textbf{0.66} & 0.74 \\ 
 & 3 & 1838.98 & 0.13 & 0.16 & \textbf{0.02} & \textbf{0.02} \\ 
 & 4 & 2454.44 & 0.51 & 0.28 & 0.42 & \textbf{0.22} \\ 
\hline 
 \multirow{3}{*}{w-P400D10$\%\bar{D}$90} & 2 & 2463.59 & 1.05 & 0.69 & 0.62 & \textbf{0.48} \\ 
 & 3 & 3688.59 & 0.10 & 0.13 & \textbf{0.01} & \textbf{0.01} \\ 
 & 4 & 4922.46 & 0.56 & 0.26 & 0.30 & \textbf{0.17} \\ 
\hline 
\hline 
\multicolumn{3}{ r|}{Average} & 0.47 & 0.43 & 0.23 & \textbf{0.19} \\ 
\noalign{\smallskip}\hline

\end{tabular}
\end{table}

We can observe that the \emph{splitting hyperplane reallocation} neighborhood ($\mathcal{N}_{3}$) is an important neighborhood for improving the quality of the solutions obtained by our VNS. 
This is not surprising given that it is the only one able to change photo-processing times of sub-regions.
The best results are usually obtained when including the neighborhoods $\mathcal{N}_{1}$ and $\mathcal{N}_{2}$ before $\mathcal{N}_{3}$ (i.e., $\mathcal{N}_{1}+\mathcal{N}_{2}+\mathcal{N}_{3}$ configuration).
This happens since the $\mathcal{N}_{1}+\mathcal{N}_{2}+\mathcal{N}_{3}$ configuration adjusts both the sub-region to drone assignments and the sub-region's photo processing times.
In fact, this configuration yields the best average deviations for 57 out of 66 cases (including unweighted and weighted instances), i.e. , 86.36\% of the cases. 
Therefore, we will consider hereafter the configuration $\mathcal{N}_{1}+\mathcal{N}_{2}+\mathcal{N}_{3}$ within our VNS heuristic.

\vspace{-3mm}
\subsection{Comparison of the proposed methods}
\label{sec:methods_comparison}

This section evaluates the performance of the proposed heuristics as well as their sensitivity with respect to the reliability factor $\sigma$ and to the maximum allowed transmission time $\hat{T}$.
For the decomposition-based heuristic (named “Decomposition”), we set a time limit of 2~hours for the MIP formulation solved in each iteration and the overall execution time of the decomposition is limited to 24 hours (i.e., 86,400s). 
In contrast,  the VNS was allowed to run for 300 seconds. 

The makespan values in seconds (columns “$T_{\max}$”) obtained by the proposed methods for the unweighted and weighted instances are reported in Tables~\ref{tab:comparison_unweighted} and \ref{tab:comparison_weighted}, respectively. Results are reported for each instance for  various values of $\sigma$, ranging from 1 to $|\bar{D}|-1$ (presented under column labelled “$\sigma$”).
A trace (“$-$”) is placed whenever a method cannot find a feasible solution until before being halted. 
For this first series of experiments the communication constraints were relaxed, i.e., $\hat{T}$ is set to a value sufficiently large.
Both tables report the instance employed at each row (column “Instance”),
the optimal/best known makespan for an instance configuration (column “\textsc{opt}”), 
the CPU time (in seconds) spent by CPLEX and the decomposition heuristic when solving an instance (columns “sec.”),
the best and average makespan values found by the VNS heuristic out of 20 executions (column “Best $T_{\max}$” and “Avg$T_{\max}$”, respectively), and
the average CPU time (in seconds) spent by the VNS heuristic to reach the solution returned at the end of its execution (column “Avg.(s)”).

{
\centering
\footnotesize
\setlength{\tabcolsep}{1pt}
\renewcommand{\arraystretch}{1.2}

\begin{longtable}{ l  r r r | r r | r r r }

\caption{Makespan($T_{\max}$) in seconds when solving unweighted instances and varying $\sigma$.}
\label{tab:comparison_unweighted} \\

\hline\noalign{\smallskip}
\multicolumn{4}{ l }{ \multirow{2}{*}{parameters: $ \hat{T}=\infty$} }  
& \multicolumn{2}{  c }{ Decomposition}
& \multicolumn{3}{ c }{VNS} \\
\noalign{\smallskip}\cline{5-9}\noalign{\smallskip}
& & & \multicolumn{1}{ c }{ } & \multicolumn{2}{ c }{time limit = $86400s$}  & \multicolumn{3}{ c }{stop condition = $300s$} \\
\noalign{\smallskip}\hline\noalign{\smallskip}
Instance & $\sigma$ & $\textsc{opt}$ & \multicolumn{1}{r }{sec.} 
& \multicolumn{1}{ r}{$T_{\max}$} & \multicolumn{1}{r }{sec.} 
& \multicolumn{1}{ r}{Best $T_{\max}$} & \multicolumn{1}{r}{Avg.$T_{\max}$} & \multicolumn{1}{r}{Avg.(s)}  \\
\noalign{\smallskip}\hline\noalign{\smallskip}
\endfirsthead

\hline\noalign{\smallskip}
\multicolumn{4}{ l }{ \multirow{2}{*}{parameters: $ \hat{T}=\infty$} }  
& \multicolumn{2}{  c }{ Decomposition}
& \multicolumn{3}{ c }{VNS} \\
\noalign{\smallskip}\cline{5-9}\noalign{\smallskip}
& & & \multicolumn{1}{ c }{ } & \multicolumn{2}{ c }{time limit = $86400s$}  & \multicolumn{3}{ c }{stop condition = $300s$} \\
\noalign{\smallskip}\hline\noalign{\smallskip}
Instance & $\sigma$ & $\textsc{opt}$ & \multicolumn{1}{r }{sec.} 
& \multicolumn{1}{ r}{$T_{\max}$} & \multicolumn{1}{r }{sec.} 
& \multicolumn{1}{ r}{Best $T_{\max}$} & \multicolumn{1}{r}{Avg.$T_{\max}$} & \multicolumn{1}{r}{Avg.(s)}  \\
\noalign{\smallskip}\hline\noalign{\smallskip}
\endhead

\noalign{\smallskip}\hline\noalign{\smallskip} 
\multicolumn{9}{ r }{{The table continues on next page}} \\ 
\noalign{\smallskip}\hline
\endfoot

\endlastfoot

 \multirow{2}{*}{u-P200D5$\%\bar{D}$70} & 1 & 1870.40 & 15.05 & 1870.40 & 0.26 & 1870.40 & 1870.40 & 0.01 \\ 
 & 2 & 3607.20 & 112.83 & 3607.20 & 0.31 & 3607.20 & 3607.20 & 0.01 \\ 
\hline 
 \multirow{2}{*}{u-P200D7$\%\bar{D}$50} & 1 & 1870.40 & 10.25 & 1870.40 & 0.42 & 1870.40 & 1870.40 & 0.01 \\ 
 & 2 & 3607.20 & 100.99 & 3607.20 & 0.59 & 3607.20 & 3607.20 & 0.01 \\ 
\hline 
 \multirow{2}{*}{u-P400D5$\%\bar{D}$70} & 1 & 3607.20 & 41.22 & 3607.20 & 1.03 & 3607.20 & 3607.20 & 0.01 \\ 
 & 2 & 7214.40 & 539.52 & 7214.40 & 1.14 & 7214.40 & 7214.40 & 0.02 \\ 
\hline 
 \multirow{2}{*}{u-P400D7$\%\bar{D}$50} & 1 & 3607.20 & 68.28 & 3607.20 & 1.50 & 3607.20 & 3607.20 & 0.02 \\ 
 & 2 & 7214.40 & 1637.54 & 7214.40 & 1.54 & 7214.40 & 7214.40 & 0.03 \\ 
\hline 
\hline 
 \multirow{3}{*}{u-P200D5$\%\bar{D}$90} & 1 & 1336.00 & 7.06 & 1336.00 & 0.10 & 1336.00 & 1336.00 & 0.01 \\ 
 & 2 & 2672.00 & 1680.46 & 2672.00 & 0.12 & 2672.00 & 2672.00 & 0.01 \\ 
 & 3 & 4008.00 & 870.76 & 4008.00 & 0.11 & 4008.00 & 4008.00 & 0.50 \\ 
\hline 
 \multirow{3}{*}{u-P200D7$\%\bar{D}$70} & 1 & 1336.00 & 6.41 & 1336.00 & 0.19 & 1336.00 & 1336.00 & 0.04 \\ 
 & 2 & 2672.00 & 2296.34 & 2672.00 & 0.26 & 2672.00 & 2672.00 & 0.01 \\ 
 & 3 & 4008.00 & 1150.97 & 4008.00 & 0.21 & 4008.00 & 4008.00 & 0.71 \\ 
\hline 
 \multirow{3}{*}{u-P400D5$\%\bar{D}$90} & 1 & 2672.00 & 35.02 & 2672.00 & 0.47 & 2672.00 & 2672.00 & 0.04 \\ 
 & 2 & *5344.00 & 86400.00 & 5344.00 & 0.44 & 5344.00 & 5344.00 & 0.01 \\ 
 & 3 & 8016.00 & 4593.69 & 8016.00 & 0.46 & 8016.00 & 8016.00 & 4.66 \\ 
\hline 
 \multirow{3}{*}{u-P400D7$\%\bar{D}$70} & 1 & 2672.00 & 59.71 & 2672.00 & 0.58 & 2672.00 & 2672.00 & 0.12 \\ 
 & 2 & *5344.00 & 86400.00 & 5344.00 & 0.76 & 5344.00 & 5344.00 & 0.01 \\ 
 & 3 & 8016.00 & 5954.75 & 8016.00 & 0.79 & 8016.00 & 8016.00 & 6.29 \\ 
\hline 
\hline 
 \multirow{4}{*}{u-P200D10$\%\bar{D}$50} & 1 & 1068.80 & 38.94 & 1068.80 & 1.46 & 1068.80 & 1068.80 & 0.08 \\ 
 & 2 & 2137.60 & 20602.10 & 2137.60 & 0.71 & 2137.60 & 2137.60 & 0.15 \\ 
 & 3 & *3286.56 & 86400.00 & \textbf{3206.40} & 1.79 & \textbf{3206.40} & \textbf{3206.40} & 0.17 \\ 
 & 4 & 4275.20 & 56226.05 & 4275.20 & 8.87 & 4275.20 & 4275.20 & 2.21 \\ 
\hline 
 \multirow{4}{*}{u-P400D10$\%\bar{D}$50} & 1 & 2137.60 & 107.68 & 2137.60 & 5.75 & 2137.60 & 2137.60 & 0.14 \\ 
 & 2 & *4275.20 & 86400.00 & 4275.20 & 2.87 & 4275.20 & 4275.20 & 0.24 \\ 
 & 3 & *6519.68 & 86400.00 & \textbf{6412.80} & 3.64 & \textbf{6412.80} & \textbf{6412.80} & 0.90 \\ 
 & 4 & *8550.40 & 86400.00 & 8550.40 & 2.93 & 8550.40 & 8550.40 & 5.27 \\ 
\hline 
\hline 
 \multirow{5}{*}{u-P200D7$\%\bar{D}$90} & 1 & 908.48 & 49.52 & 908.48 & 8.12 & 908.48 & 908.48 & 1.73 \\ 
 & 2 & *1816.96 & 86400.00 & \textbf{1790.24} & 17.56 & \textbf{1790.24} & \textbf{1790.24} & 0.12 \\ 
 & 3 & *2672.00 & 86400.00 & 2672.00 & 14.76 & 2672.00 & 2672.00 & 0.01 \\ 
 & 4 & *3607.20 & 86400.00 & \textbf{3580.48} & 23.89 & \textbf{3580.48} & \textbf{3580.48} & 0.46 \\ 
 & 5 & *4542.40 & 86400.00 & \textbf{4488.96} & 444.98 & \textbf{4488.96} & \textbf{4526.37} & 21.96 \\ 
\hline 
 \multirow{5}{*}{u-P400D7$\%\bar{D}$90} & 1 & *1816.96 & 86400.00 & 1843.68 & 13.32 & 1816.96 & 1816.96 & 1.11 \\ 
 & 2 & *3580.48 & 86400.00 & 3580.48 & 14.69 & 3580.48 & 3580.48 & 0.31 \\ 
 & 3 & *5477.60 & 86400.00 & \textbf{5344.00} & 15.52 & \textbf{5344.00} & \textbf{5344.00} & 0.01 \\ 
 & 4 & *7748.80 & 86400.00 & \textbf{7160.96} & 2449.49 & \textbf{7134.24} & \textbf{7134.24} & 8.05 \\ 
 & 5 & *10688.00 & 86400.00 & \textbf{8924.48} & 64.40 & \textbf{8924.48} & \textbf{9018.00} & 63.95 \\ 
\hline 
\hline 
 \multirow{6}{*}{u-P200D10$\%\bar{D}$70} & 1 & *801.60 & 86400.00 & 801.60 & 108.54 & 801.60 & 801.60 & 1.85 \\ 
 & 2 & *1603.20 & 86400.00 & \textbf{1549.76} & 1136.86 & \textbf{1549.76} & \textbf{1549.76} & 10.43 \\ 
 & 3 & - & - & 2297.92 & 30.52 & 2297.92 & 2299.26 & 74.77 \\ 
 & 4 & - & - & 3072.80 & 55.08 & 3072.80 & 3074.14 & 40.66 \\ 
 & 5 & *4328.64 & 86400.00 & \textbf{3847.68} & 597.09 & \textbf{3847.68} & \textbf{3847.68} & 17.27 \\ 
 & 6 & *5718.08 & 86400.00 & \textbf{4595.84} & 184.89 & \textbf{4649.28} & \textbf{4670.66} & 51.14 \\ 
\hline 
 \multirow{6}{*}{u-P400D10$\%\bar{D}$70} & 1 & *1603.20 & 86400.00 & \textbf{1549.76} & 2183.17 & \textbf{1549.76} & \textbf{1563.12} & 64.66 \\ 
 & 2 & - & - & 3072.80 & 16065.91 & 3072.80 & 3072.80 & 56.34 \\ 
 & 3 & - & - & 4595.84 & 296.30 & 4595.84 & 4602.52 & 70.87 \\ 
 & 4 & - & - & 6118.88 & 7915.51 & 6118.88 & 6137.58 & 34.18 \\ 
 & 5 & - & - & 7668.64 & 21612.06 & 7641.92 & 7686.01 & 62.77 \\ 
 & 6 & - & - & 9191.68 & 5877.11 & 9218.40 & 9352.00 & 79.74 \\ 
\hline 
 \multirow{6}{*}{u-P500D15$\%\bar{D}$50} & 1 & 1923.84 & 16185.19 & 1923.84 & 329.20 & 1923.84 & 1959.91 & 97.38 \\ 
 & 2 & - & - & 3847.68 & 14417.75 & 3847.68 & 3847.68 & 13.77 \\ 
 & 3 & - & - & 5744.80 & 21621.88 & 5744.80 & 5747.47 & 76.45 \\ 
 & 4 & - & - & 7641.92 & 14412.51 & 7641.92 & 7688.68 & 36.03 \\ 
 & 5 & - & - & 9565.76 & 14413.03 & 9565.76 & 9611.18 & 27.35 \\ 
 & 6 & - & - & 11489.60 & 8106.20 & 11489.60 & 11696.68 & 58.85 \\ 
\hline 
 \multirow{6}{*}{u-P750D15$\%\bar{D}$50} & 1 & *3126.24 & 86400.00 & \textbf{2885.76} & 270.72 & \textbf{2885.76} & \textbf{2933.86} & 52.65 \\ 
 & 2 & - & - & 5744.80 & 17315.08 & 5744.80 & 5771.52 & 78.88 \\ 
 & 3 & - & - & 8657.28 & 14419.11 & 8630.56 & 8649.26 & 91.33 \\ 
 & 4 & - & - & 11462.88 & 14419.09 & 11489.60 & 11508.30 & 76.75 \\ 
 & 5 & - & - & 14348.64 & 14421.91 & 14428.80 & 14434.14 & 35.91 \\ 
 & 6 & - & - & 17234.40 & 14418.32 & 17234.40 & 17485.57 & 37.64 \\ 
\hline 
 \multirow{6}{*}{u-P1000D15$\%\bar{D}$50} & 1 & *3847.68 & 86400.00 & 3847.68 & 7424.05 & 3847.68 & 3950.55 & 64.74 \\ 
 & 2 & - & - & 7695.36 & 14439.34 & 7641.92 & 7682.00 & 36.38 \\ 
 & 3 & - & - & - & 86400.00 & 11489.60 & 11605.83 & 67.56 \\ 
 & 4 & - & - & - & 86400.00 & 15283.84 & 15349.30 & 76.41 \\ 
 & 5 & - & - & 19131.52 & 14433.69 & 19104.80 & 19234.39 & 66.49 \\ 
 & 6 & - & - & - & 86400.00 & 22979.20 & 23410.73 & 112.49 \\ 
\hline 
\hline 
 \multirow{8}{*}{u-P200D10$\%\bar{D}$90} & 1 & *641.28 & 86400.00 & 641.28 & 3201.67 & 641.28 & 641.28 & 17.94 \\ 
 & 2 & - & - & 1202.40 & 3589.18 & 1202.40 & 1203.74 & 10.09 \\ 
 & 3 & *2030.72 & 86400.00 & \textbf{1790.24} & 297.44 & \textbf{1790.24} & \textbf{1790.24} & 0.15 \\ 
 & 4 & - & - & 2404.80 & 10993.74 & 2378.08 & 2400.79 & 1.89 \\ 
 & 5 & - & - & 2992.64 & 1200.84 & 2992.64 & 2992.64 & 12.98 \\ 
 & 6 & - & - & 3580.48 & 7224.15 & 3580.48 & 3580.48 & 0.21 \\ 
 & 7 & - & - & 4168.32 & 2382.04 & 4168.32 & 4217.75 & 43.90 \\ 
 & 8 & - & - & 4782.88 & 7820.39 & 4809.60 & 4817.62 & 13.23 \\ 
\hline 
 \multirow{8}{*}{u-P400D10$\%\bar{D}$90} & 1 & 1202.40 & 17470.88 & 1202.40 & 3376.48 & 1202.40 & 1241.14 & 131.36 \\ 
 & 2 & - & - & 2378.08 & 903.16 & 2404.80 & 2404.80 & 8.97 \\ 
 & 3 & - & - & 3580.48 & 262.56 & 3580.48 & 3580.48 & 0.08 \\ 
 & 4 & - & - & 4756.16 & 991.95 & 4756.16 & 4777.54 & 58.81 \\ 
 & 5 & - & - & 5958.56 & 255.96 & 5958.56 & 5975.93 & 59.92 \\ 
 & 6 & - & - & 7134.24 & 862.01 & 7134.24 & 7134.24 & 1.96 \\ 
 & 7 & - & - & 8336.64 & 183.60 & 8336.64 & 8379.39 & 69.29 \\ 
 & 8 & - & - & 9512.32 & 240.45 & 9619.20 & 9619.20 & 31.49 \\ 
\noalign{\smallskip}\hline
	
\end{longtable}
}

{
\centering
\footnotesize
\setlength{\tabcolsep}{1pt}
\renewcommand{\arraystretch}{1.2}

\begin{longtable}{ l  r r r | r r | r r r }

\caption{Makespan($T_{\max}$) in seconds when solving weighted instances and varying $\sigma$.}
\label{tab:comparison_weighted} \\

\hline\noalign{\smallskip}
\multicolumn{4}{ l }{ \multirow{2}{*}{parameters: $ \hat{T}=\infty$} }  
& \multicolumn{2}{  c }{ Decomposition}
& \multicolumn{3}{ c }{VNS} \\
\noalign{\smallskip}\cline{5-9}\noalign{\smallskip}
& & & \multicolumn{1}{ c }{ } & \multicolumn{2}{ c }{time limit = $86400s$}  & \multicolumn{3}{ c }{stop condition = $300s$} \\
\noalign{\smallskip}\hline\noalign{\smallskip}
Instance & $\sigma$ & $\textsc{opt}$ & \multicolumn{1}{r }{sec.} 
& \multicolumn{1}{ r}{$T_{\max}$} & \multicolumn{1}{r }{sec.} 
& \multicolumn{1}{ r}{Best $T_{\max}$} & \multicolumn{1}{r}{Avg.$T_{\max}$} & \multicolumn{1}{r}{Avg.(s)}  \\
\noalign{\smallskip}\hline\noalign{\smallskip}
\endfirsthead

\hline\noalign{\smallskip}
\multicolumn{4}{ l }{ \multirow{2}{*}{parameters: $ \hat{T}=\infty$} }  
& \multicolumn{2}{  c }{ Decomposition}
& \multicolumn{3}{ c }{VNS} \\
\noalign{\smallskip}\cline{5-9}\noalign{\smallskip}
& & & \multicolumn{1}{ c }{ } & \multicolumn{2}{ c }{time limit = $86400s$}  & \multicolumn{3}{ c }{stop condition = $300s$} \\
\noalign{\smallskip}\hline\noalign{\smallskip}
Instance & $\sigma$ & $\textsc{opt}$ & \multicolumn{1}{r }{sec.} 
& \multicolumn{1}{ r}{$T_{\max}$} & \multicolumn{1}{r }{sec.} 
& \multicolumn{1}{ r}{Best $T_{\max}$} & \multicolumn{1}{r}{Avg.$T_{\max}$} & \multicolumn{1}{r}{Avg.(s)}  \\
\noalign{\smallskip}\hline\noalign{\smallskip}
\endhead

\noalign{\smallskip}\hline \noalign{\smallskip}
\multicolumn{9}{ r }{{The table continues on next page}} \\ 
\noalign{\smallskip}\hline
\endfoot

\endlastfoot

 \multirow{2}{*}{w-P200D5$\%\bar{D}$70} & 1 & 1886.98 & 15.84 & 1886.98 & 0.39 & 1886.98 & 1886.98 & 0.01 \\ 
 & 2 & 3699.67 & 172.46 & 3699.67 & 0.37 & 3699.67 & 3699.67 & 0.02 \\ 
\hline 
 \multirow{2}{*}{w-P200D7$\%\bar{D}$50} & 1 & 1885.31 & 22.74 & 1946.84 & 0.41 & 1885.31 & 1885.31 & 0.03 \\ 
 & 2 & 3740.19 & 653.34 & 3740.19 & 0.47 & 3740.19 & 3740.19 & 0.04 \\ 
\hline 
 \multirow{2}{*}{w-P400D5$\%\bar{D}$70} & 1 & 3735.32 & 79.99 & 3735.32 & 1.20 & 3735.32 & 3735.32 & 0.01 \\ 
  & 2 & 7457.00 & 1988.79 & 7458.55 & 1.22 & 7458.55 & 7458.55 & 0.19 \\ 
\hline 
 \multirow{2}{*}{w-P400D7$\%\bar{D}$50} & 1 & 3773.12 & 94.42 & 3773.12 & 2.47 & 3773.12 & 3773.12 & 0.03 \\ 
 & 2 & 7436.46 & 1464.54 & 7436.46 & 1.53 & 7436.46 & 7436.46 & 0.26 \\ 
\hline 
\hline 
 \multirow{3}{*}{w-P200D5$\%\bar{D}$90} & 1 & 1409.48 & 212.44 & 1409.48 & 0.26 & 1409.48 & 1409.48 & 0.08 \\ 
 & 2 & 2735.45 & 34294.09 & 2736.79 & 0.33 & 2735.45 & 2736.23 & 0.90 \\ 
 & 3 & *4148.52 & 86400.00 & 4148.52 & 0.34 & 4148.52 & 4148.52 & 0.78 \\ 
\hline 
 \multirow{3}{*}{w-P200D7$\%\bar{D}$70} & 1 & 1399.82 & 453.18 & 1399.82 & 0.55 & 1399.82 & 1399.82 & 0.10 \\ 
 & 2 & 2724.20 & 2647.32 & 2726.87 & 0.50 & 2724.20 & 2725.04 & 0.40 \\ 
 & 3 & *4105.25 & 86400.00 & 4105.25 & 0.59 & 4105.25 & 4105.25 & 3.41 \\ 
\hline 
 \multirow{3}{*}{w-P400D5$\%\bar{D}$90} & 1 & 2812.93 & 925.85 & 2812.93 & 0.56 & 2812.93 & 2812.93 & 0.14 \\ 
 & 2 & *5555.98 & 86400.00 & 5568.43 & 0.75 & \textbf{5555.12} & \textbf{5555.30} & 13.20 \\ 
 & 3 & *8381.36 & 86400.00 & 8381.36 & 0.74 & 8381.36 & 8381.36 & 4.07 \\ 
\hline 
 \multirow{3}{*}{w-P400D7$\%\bar{D}$70} & 1 & 2791.66 & 605.48 & 2791.66 & 1.74 & 2791.66 & 2791.66 & 0.15 \\ 
 & 2 & *5545.75 & 86400.00 & 5555.17 & 1.37 & \textbf{5545.49} & \textbf{5545.51} & 3.73 \\ 
 & 3 & *8346.83 & 86400.00 & 8346.83 & 1.43 & 8346.83 & 8346.83 & 13.77 \\ 
\hline 
\hline 
 \multirow{4}{*}{w-P200D10$\%\bar{D}$50} & 1 & 1121.89 & 3611.52 & 1121.89 & 13.61 & 1121.89 & 1121.89 & 42.04 \\ 
 & 2 & *2222.97 & 86400.00 & \textbf{2216.51} & 99.52 & \textbf{2212.64} & \textbf{2213.97} & 126.59 \\ 
 & 3 & *3329.75 & 86400.00 & \textbf{3314.91} & 69.04 & \textbf{3314.91} & \textbf{3320.12} & 54.75 \\ 
 & 4 & *4509.50 & 86400.00 & \textbf{4436.74} & 19.78 & \textbf{4436.74} & \textbf{4437.46} & 76.78 \\ 
\hline 
 \multirow{4}{*}{w-P400D10$\%\bar{D}$50} & 1 & 2290.48 & 40443.02 & 2316.78 & 34.98 & 2290.48 & 2290.48 & 6.50 \\ 
 & 2 & *4591.18 & 86400.00 & \textbf{4445.09} & 16.37 & \textbf{4434.52} & \textbf{4437.33} & 78.96 \\ 
 & 3 & *6765.02 & 86400.00 & \textbf{6650.29} & 19.40 & \textbf{6648.11} & \textbf{6657.81} & 86.10 \\ 
 & 4 & *9938.25 & 86400.00 & \textbf{8901.19} & 73.87 & \textbf{8901.19} & \textbf{8902.89} & 58.79 \\ 
\hline 
\hline 
 \multirow{5}{*}{w-P200D7$\%\bar{D}$90} & 1 & 944.06 & 8669.90 & 944.06 & 17.64 & 944.06 & 944.06 & 58.91 \\ 
 & 2 & *1838.27 & 86400.00 & \textbf{1818.19} & 315.24 & \textbf{1816.30} & \textbf{1816.84} & 94.62 \\ 
 & 3 & *2737.73 & 86400.00 & \textbf{2724.26} & 193.49 & \textbf{2724.19} & \textbf{2724.19} & 151.52 \\ 
 & 4 & *3784.47 & 86400.00 & \textbf{3633.31} & 2826.81 & \textbf{3632.45} & \textbf{3633.19} & 149.25 \\ 
 & 5 & *4952.71 & 86400.00 & \textbf{4582.94} & 2834.41 & \textbf{4585.65} & \textbf{4600.59} & 133.66 \\ 
\hline 
 \multirow{5}{*}{w-P400D7$\%\bar{D}$90} & 1 & *1909.18 & 86400.00 & \textbf{1891.23} & 54.26 & \textbf{1891.23} & \textbf{1904.24} & 125.80 \\ 
 & 2 & *3922.21 & 86400.00 & \textbf{3701.03} & 319.12 & \textbf{3697.34} & \textbf{3697.90} & 113.25 \\ 
 & 3 & *5567.01 & 86400.00 & \textbf{5545.72} & 49.86 & \textbf{5545.48} & \textbf{5545.48} & 78.77 \\ 
 & 4 & *9094.33 & 86400.00 & \textbf{7398.95} & 7395.50 & \textbf{7394.18} & \textbf{7396.18} & 130.84 \\ 
 & 5 & - & - & 9302.92 & 15918.34 & 9300.46 & 9361.46 & 113.56 \\ 
\hline 
\hline 
 \multirow{6}{*}{w-P200D10$\%\bar{D}$70} & 1 & 820.40 & 62142.01 & 822.78 & 1015.08 & 825.06 & 825.06 & 81.79 \\ 
 & 2 & - & - & 1577.97 & 11209.34 & 1577.87 & 1582.31 & 115.51 \\ 
 & 3 & *2446.63 & 86400.00 & \textbf{2372.45} & 14413.80 & \textbf{2365.78} & \textbf{2371.93} & 149.33 \\ 
 & 4 & *4168.27 & 86400.00 & \textbf{3159.91} & 8257.01 & \textbf{3157.05} & \textbf{3165.93} & 107.74 \\ 
 & 5 & *4198.12 & 86400.00 & \textbf{3945.45} & 22388.35 & \textbf{3947.86} & \textbf{3958.33} & 68.83 \\ 
 & 6 & *5613.12 & 86400.00 & \textbf{4763.35} & 5832.36 & \textbf{4763.35} & \textbf{4809.49} & 58.26 \\ 
\hline 
 \multirow{6}{*}{w-P400D10$\%\bar{D}$70} & 1 & *1671.12 & 86400.00 & \textbf{1632.77} & 7249.23 & \textbf{1633.77} & \textbf{1639.62} & 114.80 \\ 
 & 2 & *3303.16 & 86400.00 & \textbf{3166.70} & 27629.18 & \textbf{3164.05} & \textbf{3175.29} & 134.49 \\ 
 & 3 & - & - & 4748.65 & 21612.15 & 4747.60 & 4756.31 & 127.08 \\ 
 & 4 & *7080.11 & 86400.00 & \textbf{6335.64} & 28821.90 & \textbf{6334.85} & \textbf{6347.79} & 101.82 \\ 
  & 5 & *8989.29 & 86400.00 & \textbf{7918.13} & 21613.88 & \textbf{7911.42} & \textbf{7943.12} & 96.72 \\ 
 & 6 & - & - & 9529.57 & 21616.77 & 9534.57 & 9630.75 & 73.83 \\ 
\hline 
 \multirow{6}{*}{w-P500D15$\%\bar{D}$50} & 1 & *2028.11 & 86400.00 & \textbf{1979.39} & 1826.05 & \textbf{1988.39} & \textbf{2007.95} & 55.76 \\ 
 & 2 & - & - & 3905.71 & 11820.59 & 3902.44 & 3913.18 & 128.80 \\ 
 & 3 & - & - & 5851.96 & 21620.22 & 5855.56 & 5864.11 & 144.98 \\ 
 & 4 & - & - & 7800.26 & 14410.70 & 7804.04 & 7818.67 & 135.90 \\ 
 & 5 & - & - & 9753.65 & 28826.40 & 9750.66 & 9790.69 & 95.56 \\ 
 & 6 & - & - & 11730.98 & 21619.57 & 11776.53 & 11904.97 & 112.61 \\ 
\hline 
 \multirow{6}{*}{w-P750D15$\%\bar{D}$50} & 1 & *3138.73 & 86400.00 & \textbf{2984.54} & 2474.51 & \textbf{3008.45} & \textbf{3047.97} & 105.02 \\ 
 & 2 & - & - & 5849.27 & 7556.80 & 5844.06 & 5858.75 & 101.48 \\ 
 & 3 & - & - & 8767.62 & 19072.32 & 8762.21 & 8784.29 & 126.76 \\ 
 & 4 & - & - & 11680.18 & 21624.97 & 11679.59 & 11712.72 & 144.81 \\ 
 & 5 & - & - & 14602.86 & 19282.36 & 14610.90 & 14665.37 & 80.64 \\ 
 & 6 & - & - & - & 86400.00 & 17665.34 & 17820.95 & 130.71 \\ 
\hline 
 \multirow{6}{*}{w-P1000D15$\%\bar{D}$50} & 1 & *4343.26 & 86400.00 & \textbf{3917.65} & 7657.63 & \textbf{3962.55} & \textbf{4020.12} & 86.24 \\ 
 & 2 & - & - & 7740.00 & 14407.20 & 7724.25 & 7752.53 & 142.33 \\ 
 & 3 & - & - & - & 86400.00 & 11594.43 & 11614.93 & 122.14 \\ 
 & 4 & - & - & - & 86400.00 & 15467.11 & 15493.21 & 112.32 \\ 
 & 5 & - & - & 19306.31 & 11764.35 & 19315.50 & 19383.26 & 112.98 \\ 
 & 6 & - & - & 23253.94 & 14438.73 & 23309.74 & 23647.27 & 108.75 \\ 
\hline 
\hline 
 \multirow{8}{*}{w-P200D10$\%\bar{D}$90} & 1 & *665.54 & 86400.00 & \textbf{646.58} & 15579.50 & \textbf{659.08} & \textbf{662.94} & 121.04 \\ 
 & 2 & - & - & 1237.80 & 21616.32 & 1230.42 & 1239.56 & 74.07 \\ 
 & 3 & - & - & 1840.77 & 28840.56 & 1838.98 & 1839.37 & 104.40 \\ 
 & 4 & - & - & 2460.09 & 21622.26 & 2454.44 & 2459.90 & 81.69 \\ 
 & 5 & - & - & 3069.42 & 28825.47 & 3067.49 & 3076.62 & 90.21 \\ 
 & 6 & - & - & 3678.99 & 36027.84 & 3678.08 & 3678.69 & 92.09 \\ 
 & 7 & - & - & 4305.23 & 21622.02 & 4300.25 & 4330.28 & 105.68 \\ 
 & 8 & - & - & 4943.87 & 21621.77 & 4954.66 & 4978.82 & 73.89 \\ 
\hline 
 \multirow{8}{*}{w-P400D10$\%\bar{D}$90} & 1 & *1302.82 & 86400.00 & \textbf{1265.11} & 9770.29 & \textbf{1270.19} & 1309.43 & 105.74 \\ 
 & 2 & - & - & 2468.59 & 14732.29 & 2467.42 & 2475.52 & 89.36 \\ 
 & 3 & - & - & 3690.58 & 21646.61 & 3688.64 & 3689.12 & 145.47 \\ 
 & 4 & - & - & 4925.60 & 21659.91 & 4924.38 & 4930.98 & 105.31 \\ 
 & 5 & - & - & 6151.24 & 21651.88 & 6158.54 & 6169.61 & 98.90 \\ 
 & 6 & - & - & 7378.95 & 21647.51 & 7377.22 & 7378.39 & 89.97 \\ 
 & 7 & - & - & 8615.73 & 19407.21 & 8611.91 & 8655.26 & 101.35 \\ 
 & 8 & - & - & 9893.10 & 17639.44 & 9929.49 & 9981.36 & 75.65 \\ 
\noalign{\smallskip}\hline	
	
\end{longtable}
}

We observe from the tables that both decomposition and VNS methods performed well for all instances tested.
In fact, both methods find solutions when the exact method fails. 
In general, it appears that increasing the reliability factor does not affect the performance of the proposed heuristic methods.

Furthermore, the decomposition could reach the optimum or improve (values indicated in bold) the best known makespan (cases with “$\ast$” in the column “\textsc{opt}”) in several instances, except for the cases 
“u-P400D7$\%D90$” ($\sigma=1$),  
“w-P200D7$\%D50$” ($\sigma=1$),  
“w-P400D5$\%\bar{D}$70” ($\sigma=2$),  
“w-P200D5$\%\bar{D}$90” ($\sigma=2$),  
“w-P400D5$\%\bar{D}$90” ($\sigma=2$), 
“w-P400D10$\%D50$” ($\sigma=1$),  \\
“w-P200D7$\%\bar{D}$70”($\sigma=2$),  
“w-P400D7$\%\bar{D}$70”($\sigma=2$), and 
“w-P200D10$\%D70$” ($\sigma=1$)
for which more iterations without improving the objective function need to be done by the method in order to be able to obtain best known solutions (as discussed in Section \ref{sec:decomp_vs_rcapsac}).
Otherwise, it kept the percentage deviations w.r.t. the optimum below 1.47\% (“u-P400D7$\%\bar{D}$90 and $\sigma = 1$”) for the unweighted instances and below 3.26\% (“w-P200D7$\%\bar{D}$50” and $\sigma = 1$) for the weighted cases. 
Moreover, the decomposition method usually requires considerable smaller execution times than the time spent by CPLEX (first column “sec”) to find the optima (or the best known solution for the cases with “$\ast$”) reported in  the column “\textsc{opt}”.

Regarding the VNS-based heuristic, it also presented good performance. It is the fastest proposed method, with average makespan deviation w.r.t. the best known solution always inferior to 3.22\%(“u-P400D10$\%\bar{D}$90” and $\sigma = 1$) for unweighted instances, and inferior to 0.57\%(“w-P200D10$\%\bar{D}$70” and $\sigma = 1$) for weighted ones. 
Besides,  VNS solutions are in average better than the best known makespan in 42 instances of the instances tested (values in bold).

Those average makespan values (column “Avg.$T_{\max}$”) improvements yield average reduction of 6.14\% and 5.17\% for the unweighted and weighted cases, respectively.
In particular, the VNS heuristic is the unique method reaching feasible solutions for the instances “u-P1000D15$\%\bar{D}$50” ($\sigma \in \{ 3,4,6 \}$), and “w-P1000D15$\%\bar{D}$50” ($\sigma \in \{ 3,4 \}$).
Finally, the VNS heuristic is more reliable in terms of finding near-optimal solutions within a small time horizon when compared to the decomposition method. 

\begin{figure}[H]
    \centering
	\includegraphics[scale=1.0]{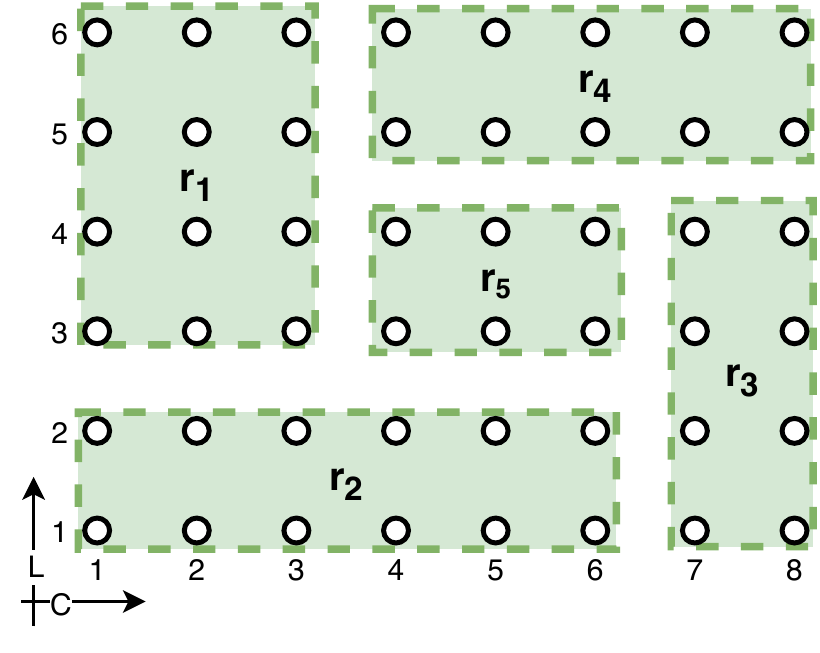}
   	\caption{Spatial-convex covering which cannot be represented by the adopted spatial-partition tree data structure.}
   	\label{fig:spatial_tree_limitation}
\end{figure}

The choice of representing a solution through a spatial partition tree as shown in Section~\ref{sec:spatial_part_tree} is very important to simplify the exploration of neighboring solutions. 
However, such representation is not able to represent all the possible spatial-convex covering of the solution space.
For instance, Figure~\ref{fig:spatial_tree_limitation} illustrates an example of covering (dashed lines) which cannot be represented by the adopted spatial-partition tree data structure. 
Representing such covering would require the use of another data structure. 
Our VNS explores rather efficiently the solution space, although it is proven not capable of visiting all the feasible solutions for the CAPsac.

Regarding the sensitivity analysis of the proposed methods to the parameter $\hat{T}$, that is done by progressively decreasing the value of $\hat{T}$ by 1 second until the communication delays turn the problem infeasible.
In particular, the experiments are concentrated upon the instances “u-P200D5$\%\bar{D}$70” and “w-P200D5$\%\bar{D}$70” with $\sigma = 1$.
In these instances, the longest communication delays found when executing the decomposition heuristic and the VNS (for 20 runs), when no constraints are imposed to the maximum transmission time (i.e. , $\hat{T} = +\infty$), vary between 33.6s and 48s. 
For that reason, the first value of $\hat{T}$ used in our experiments was 60 seconds, being decreased by 1 second for each tested $\hat{T}$  down to $\hat{T} = 23s$, when both instances are actually proved infeasible by CPLEX.

\begin{figure}[H]
    \centering
	\includegraphics[scale=0.2]{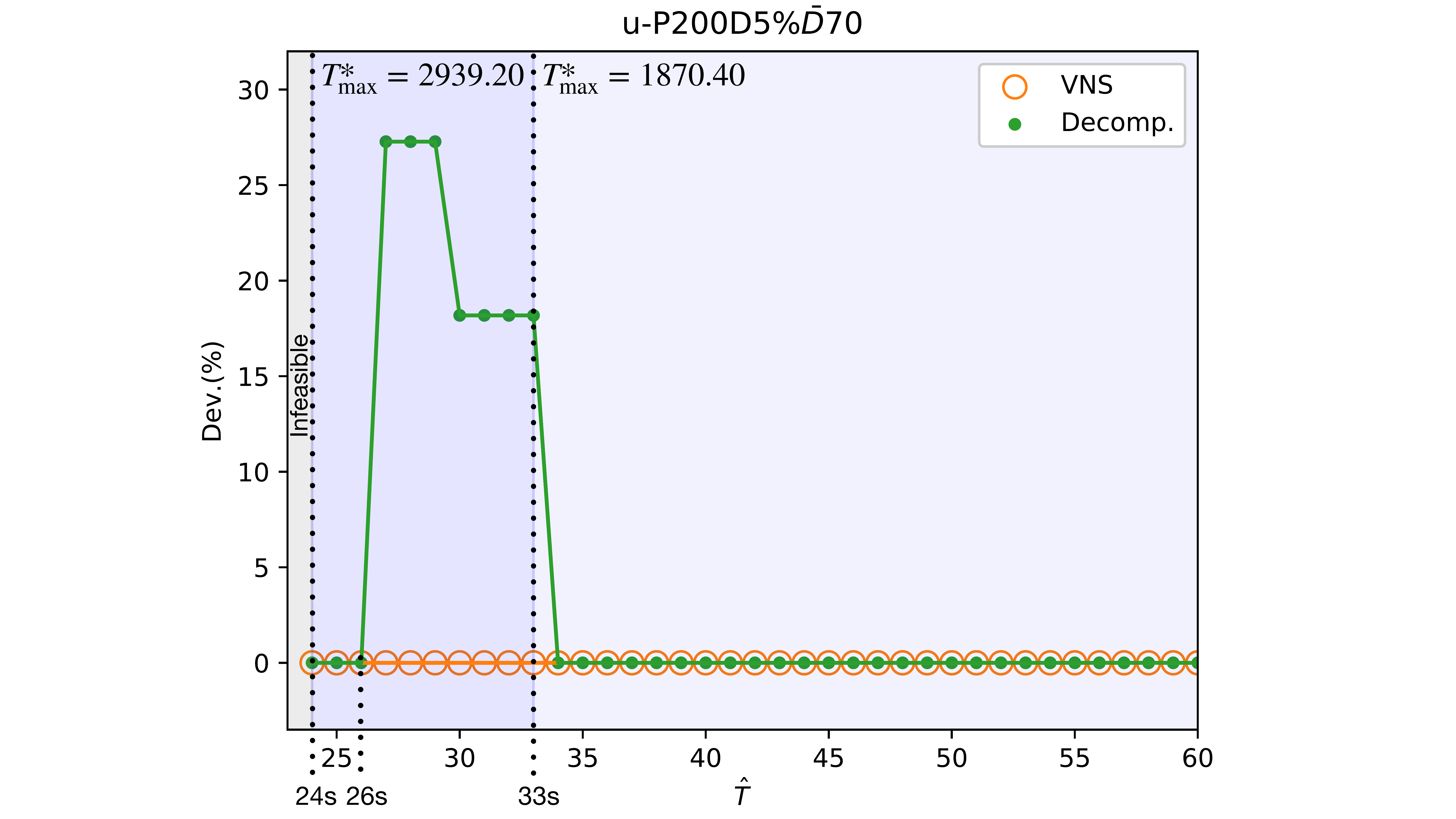}
	\includegraphics[scale=0.2]{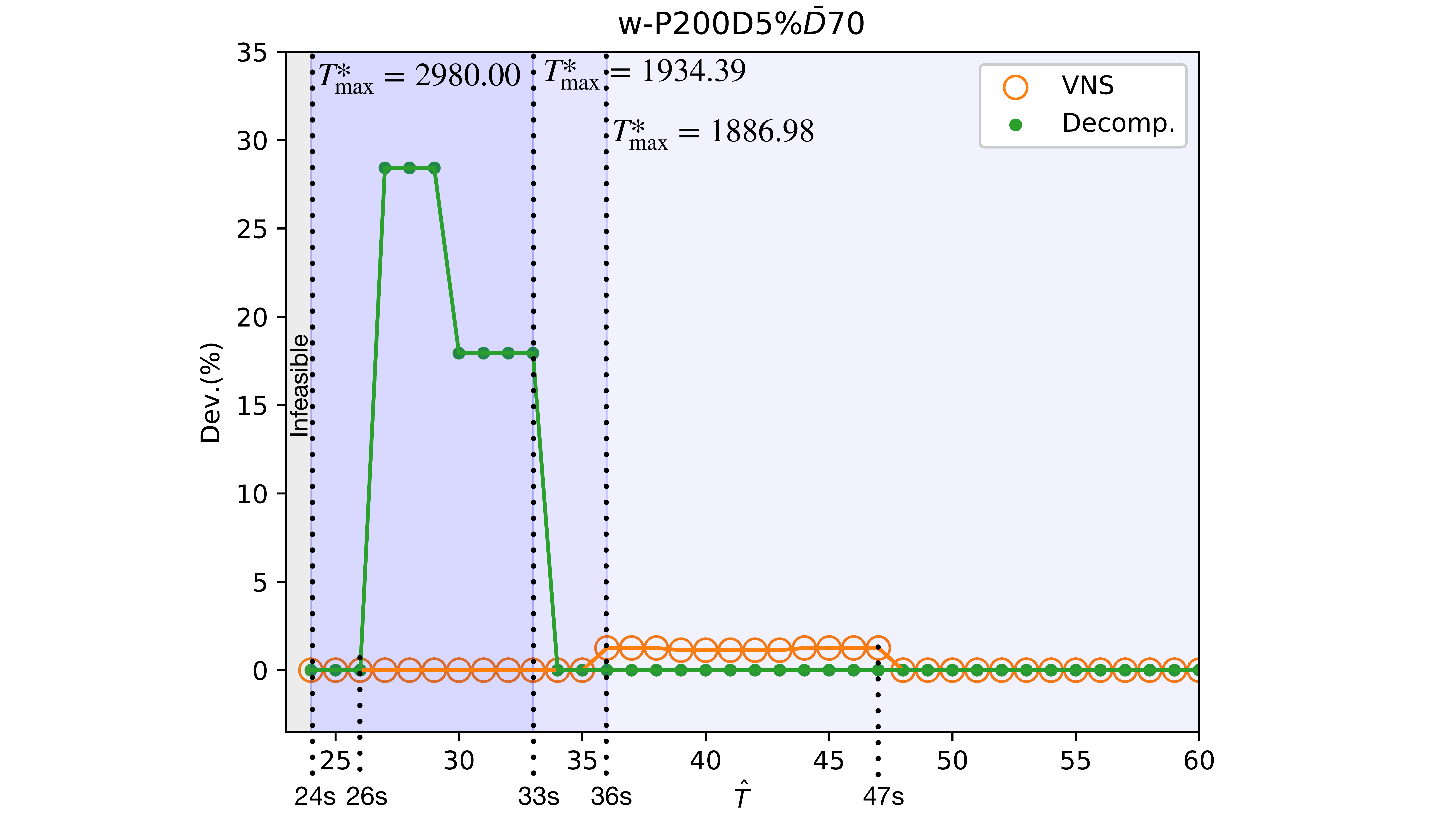}		
	\caption{Percentage deviation achieved varying $\hat{T}$ for instances “u-P200D5$\%\bar{D}$70” (top) and “w-P200D5$\%\bar{D}$70” (bottom).}	
	\label{fig:sensitivity_t_hat}
\end{figure}

Figure~\ref{fig:sensitivity_t_hat} presents the results of instances “u-P200D5$\%\bar{D}$70” and “w-P200D5$\%\bar{D}$70” obtained by the decomposition-based and the VNS heuristics for $\hat{T} \in \{24s, 25s, 26s,\ldots,59s,60s\}$.
The figure shows how decreasing $\hat{T}$ values affect the ability of the proposed methods to achieve optimal or near-optimal solutions.
We used distinct shaded regions in the figure to represent the different optimal $T_{max}^*$ values obtained within the interval of tested $\hat{T}$ values. 
Thus, we can observe for instance  “u-P200D5$\%\bar{D}$70” two shaded regions before the problem becomes infeasible for $\hat{T}<24s$, one in which $T_{\max}^* = 2939.20$ for $\hat{T} = [24s, 33s]$, and another in which $T_{\max}^* = 1870.40$ for $\hat{T} = [34s, 60s]$.
The vertical axis “Dev.($\%$)” in the figure reports the percentage deviation w.r.t. the optimum $T_{\max}^*$ in the shaded region associated with each value of $\hat{T}$ tested.
Green lines and “$\bullet$” symbols represent the deviations obtained by the decomposition-based method (named “Decomp.”) whereas the average deviation achieved by 20 runs of the VNS method (denoted by ``VNS'') are represented by the orange lines and empty “$\circ$”.

It can be observed in Figure~\ref{fig:sensitivity_t_hat} from the unweighted instance that the decomposition-based method achieves optimality for most of the tested $\hat{T}$ values except for $27s \leq \hat{T} \leq 33s$ where it reaches percentage deviations up to $27.27\%$. 
Those high deviations are mainly due to the $T_{\max}^{\ast}$ increase for $\hat{T} \leq 33s$, from 1870.40 to 2939.20. We observed that for $27s \leq \hat{T}$  the initial bounds $n_{\ell}$ and $n_{u}$ of the decomposition method are not good estimates for the cardinalities of the sub-regions that compose the optimal solutions, and thus, the heuristic is likely to finish before reaching a near-optimal solution.
For $24s \leq \hat{T} \leq 26s$, the initial limits $n_{\ell}$ and $n_{u}$ lead to infeasible problems until they are large enough to encompass feasible solutions,  which were in this case, close to optima.
In contrast, the VNS method achieves optimality for all tested $\hat{T}$ values when solving the unweighted case.

Regarding the weighted case, the decomposition method presents similar behavior as when solving the unweighted case, and its solution deviation goes up to $28.42\%$ for $27s \leq \hat{T} \leq 33s$.
The first $T_{max}^{\ast}$ increase from $1886.98$ to $1934.39$ (i.e. +2.5\%) was not sufficient to affect the quality of the solution obtained by the decomposition. 
The VNS only presents small increases in the average deviation for $36s \leq \hat{T} \leq 47s$ but they never surpass $1.25\%$, otherwise the VNS reaches optimality.

We remark that the VNS method constructs its initial solution in our experiments using $\hat{T} {=}{+}\infty$. 
This means that the VNS initial solutions may be infeasible in terms of the transmission delays. 
Our experiments allowed us to observe that the proposed VNS was able to find feasible solutions even from infeasible starting points.

\section{Conclusions}
\label{sec:conclusions}

We addressed the optimization of the 3D reconstruction step within a swarm-powered 3D mapping mission according to the so-called Covering-Assignment Problem for swarm-powered ad hoc clouds - CAPsac.
It minimizes the completion time of the 3D reconstruction photo-processing phase by exploiting the distributed computational power of the swarm of drones.
This was achieved by integrating the photo covering (workload) optimization along with the assignment of the drones to the photo processing subtasks (i.e., sub-regions). 

Since time is a very precious resource in emergency field operations supported by UAVs, the limited computation time — in the order of a couple of minutes for the biggest and more complex instances, we proposed two heuristic algorithms to solve this NP-hard problem in a limited amount of time (i) a mathematical programming-based heuristic as well as a (ii) Variable Neighborhood Search method.

Despite the fact that we had focused on a swarm-powered distributed 3D reconstruction for humanitarian emergency response application, the proposed methods can be deployed to any application in the CAPsac's scope.
Finally, the methods can be adapted to consider auxiliary computing resources in addition to those offered by the UAV swarm.

Computational experiments were conducted with the unweighted and weighted realistic instances available online at~\url{https://github.com/ds4dm/CAPsac} to assess the performance of the proposed heuristic methods.
The experiments exposed that the VNS heuristic either quickly achieves near-optimal solutions or rapidly improves the best known makespan for a vast number of instances.
The decomposition-based heuristic is also effective in most of the tested cases and is an efficient method when compared to solving the $CAPsac$ formulations by commercial solvers.
Although the performance of decomposition-based heuristic deteriorates for some values of $\hat{T}$ close to infeasibility, the sensitivity analysis done for the decomposition and VNS methods demonstrated that those methods still perform well when varying $\sigma$ and $\hat{T}$.

Our current work with Humanitas Solutions~(\url{https://www.humanitas.io/}) is focused on embedding the heuristics proposed in this paper into real deployed UAV swarms. 

\section*{Acknowledgements}
The authors would like to thank all support from Humanitas Solutions and the Canada Excellence Research Chair in Data Science for Real-Time Decision-Making. 
This work was financial supported in part by the Natural Sciences and Engineering Research Council of Canada through the Canada Excellence Research Chair in data science for real-time decision-making,  in part by Prompt, and in part by MITACS.
The computing resources were made available by Compute Canada.

\bibliographystyle{unsrtnat}
\bibliography{ms.bbl}

\end{document}